\documentclass[superscriptaddress,twocolumn,10pt,showpacs,aps,prl]{revtex4-1}
\usepackage{amsmath}
\usepackage{graphicx}
\usepackage{inputenc}
\usepackage{color}
\usepackage{setspace}
\usepackage{bm}
\usepackage{soul}
\usepackage{hyperref}

\newcommand {\bacovo} {BaCo$_2$V$_2$O$_8$}

\begin{document}

\title{
Topological quantum phase transition \\
in the Ising-like antiferromagnetic spin chain BaCo$_2$V$_2$O$_8$}

\author{Q.~Faure}
\affiliation{Universit\'e Grenoble Alpes, CEA, INAC, MEM, F-38000 Grenoble, France}

\author{S.~Takayoshi}
\affiliation{DPMC-MaNEP, University of Geneva, 24 Quai Ernest Ansermet, CH-1211 Geneva, Switzerland}

\author{S.~Petit}
\affiliation{Laboratoire L\'eon Brillouin, CEA, CNRS, Universit\'e Paris-Saclay, CE-Saclay, F-91191 Gif-sur-Yvette, France}

\author{V.~Simonet}
\email[]{virginie.simonet@neel.cnrs.fr}
\affiliation{Institut N\'eel, CNRS--UGA, F-38042 Grenoble, France}

\author{S.~Raymond}
\affiliation{Univ. Grenoble Alpes, CEA, INAC, MEM, F-38000 Grenoble, France}

\author{L.-P.~Regnault}
\affiliation{Univ. Grenoble Alpes, CEA, INAC, MEM, F-38000 Grenoble, France}

\author{M.~Boehm}
\affiliation{Institut Laue Langevin, CS 20156, F-38042 Grenoble, France}

\author{J.~S.~White}
\affiliation{Laboratory for Neutron Scattering and Imaging, PSI, CH-5232 Villigen, Switzerland}

\author{M.~M\aa nsson}
\affiliation{Materials Physics, KTH Royal Institute of Technology, Electrum 229, SE-16440 Kista, Stockholm, Sweden}

\author{Ch.~R\"uegg}
\affiliation{Laboratory for Neutron Scattering and Imaging, PSI, CH-5232 Villigen, Switzerland}

\author{P.~Lejay}
\affiliation{Institut N\'eel, CNRS--UGA, F-38042 Grenoble, France}

\author{B.~Canals}
\affiliation{Institut N\'eel, CNRS--UGA, F-38042 Grenoble, France}

\author{T.~Lorenz}
\affiliation{II. Physikalisches Institut, Universit\"at zu K\"oln, D-50937 K\"oln, Germany}

\author{S.~C.~Furuya}
\affiliation{Condensed Matter Theory Laboratory, RIKEN, Wako, Saitama 351-0198, Japan}

\author{T.~Giamarchi}
\email[]{Thierry.Giamarchi@unige.ch}
\affiliation{DPMC-MaNEP, University of Geneva, 24 Quai Ernest Ansermet, CH-1211 Geneva, Switzerland}

\author{B.~Grenier}
\affiliation{Univ. Grenoble Alpes, CEA, INAC, MEM, F-38000 Grenoble, France}

(Dated: \today)

\maketitle


{\bf
Since the seminal ideas of Berezinskii, Kosterlitz and Thouless, topological excitations are at the heart of our understanding of a whole novel class of phase transitions. In most of the cases, those transitions are controlled by a single type of topological objects. There are however some situations, still poorly understood, where two dual topological excitations fight to control the phase diagram and the transition. Finding experimental realization of such cases is thus of considerable interest. We show here that this situation occurs in BaCo$_2$V$_2$O$_8$, a spin-1/2 Ising-like quasi-one dimensional antiferromagnet when subjected to a uniform magnetic field transverse to the Ising axis. Using neutron scattering experiments, we measure a drastic modification of the quantum excitations beyond a critical value of the magnetic field. This quantum phase transition is identified, through a comparison with theoretical calculations, to be a transition between two different types of solitonic topological objects, which are captured by different components of the dynamical structure factor.}

The pioneering work of Berezinskii, Kosterlitz and Thouless (BKT)~\cite{Berezinskii,KosterlitzThouless1973} has enlightened the role played by topological excitations in the two dimensional classical $XY$ model. Since then, the topological aspects have been found to be crucial not only to a host of two dimensional classical systems \cite{BKT40years}, but also in a spectacular way in the one dimensional quantum world \cite{giamarchibook} with in particular the remarkable case of spin-1 chains~\cite{Haldane1983}. Such concepts have allowed to understand important aspects of the physics of materials such as the quantum hall effect~\cite{Thouless1982} and even predict new classes of systems such as topological insulators~\cite{Hasan2010}. Identifying and understanding the topological aspects of matter has thus become a major focus in condensed matter physics and quantum optics, where topological phases such as the Haldane model~\cite{Haldane1988} have been remarkably realized~\cite{Jotzu2014}.

For classical and quantum critical phenomena, we have by now a good understanding of the prototypical topological phase transition in which only a single topological entity controls the transition. This was the case in the original BKT work, where vortex-antivortex excitations deconfine in a similar way than electrical charges in the two dimensional Coulomb gas~\cite{Kosterlitz1974}. In the quantum world, this situation is described by the celebrated sine-Gordon model, which also plays a central role in quantum field theory \cite{rajaraman_instanton}. However, a richer and more difficult to understand class of topological transitions was rapidly pointed out to also play a major role for several systems~\cite{Jose1977,giamarchibook}. This situation arises when two conjugate fields, subjected to the Heisenberg uncertainty principle and plunged into different potentials, compete with each other. The phase diagram is thus controlled by the confinement/deconfinement of the corresponding \emph{dual} topological exc
itations. These situations are considerably more difficult to analyse \cite{Jose1977,Fertig2002} and need much more sophisticated field theory descriptions such as the so-called dual-field double sine-Gordon model \cite{GiamarchiSchulz1988,Lecheminant2002}. It is thus of considerable interest to find good experimental realizations of such cases, with the possibility to tune the system through the transition and in which the evolution of the excitations can be scrutinized.

Here we show that the Ising-like spin chain compound \bacovo\ yields a good realization of such a topological transition when subjected to a magnetic field transverse to the Ising axis (which points along the chains). This compound is characterized by unique features. First, it shows a strong Ising-like anisotropy. Next, because of the crystallographic peculiarities, applying a uniform field creates a staggered field perpendicular to both the Ising-axis and the uniform field. This specificity allows dual topological excitations to be present. Those are solitonic excitations associated with the two angles needed to parametrize a spin and which in quantum mechanics are conjugate variables. Using a combination of neutron scattering experiments, numerical calculation of the microscopic description of the system, and a field theory analysis based on the double sine-Gordon model, we show that the transition observed in \bacovo\ at a certain critical value of the magnetic field corresponds to a quantum phase transit
ion between two phases dominated by dual topological excitations (see Fig.~\ref{fig4}): i)~spinons along the chain direction --~fractionalized excitations carrying a spin 1/2~--; ii)~their dual excitations -- carrying a spin 1 -- along the axis of the staggered magnetic field.

\vspace{1\baselineskip}
{\bf \bacovo, a model system}

\begin{figure}
\begin{center}
\includegraphics[width=8.5 cm]{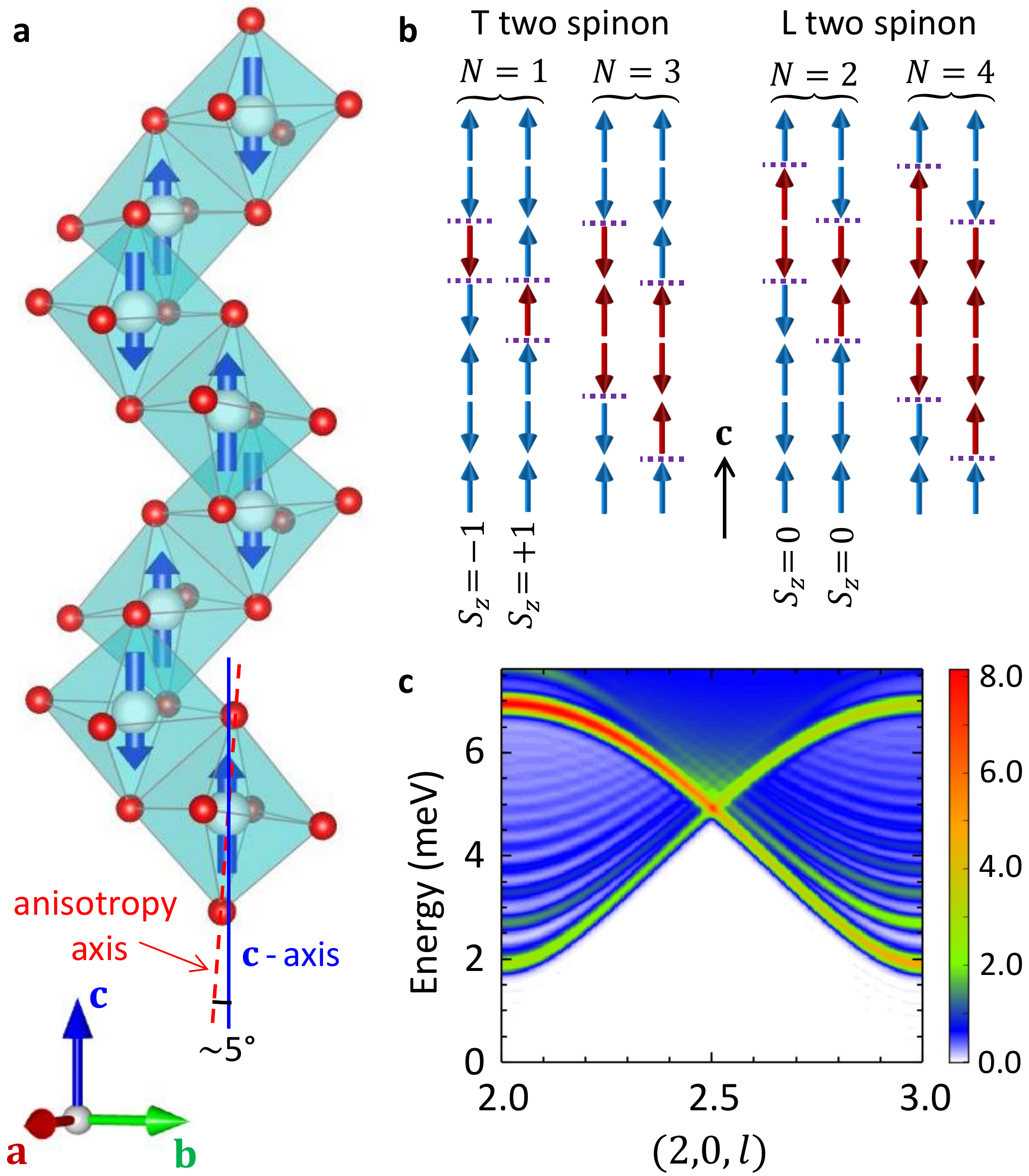}
\caption{{\bf \bacovo\ at zero field.} {\bf a}, Structure of a single Co$^{2+}$ screw chain of \bacovo\ (blue and red spheres are Co and O respectively) with the zero-field magnetic arrangement (blue arrows). The local anisotropy axis is plotted (red dashed line) for the bottom CoO$_6$ octahedron. Note that two kinds of chain parallel to $\bf c$ are equally present in \bacovo, rotating in opposite directions with respect to the $\bf c-$axis. {\bf b},~Schematic sketch of the two-spinon excitations described as domain walls in the pure Ising case. The transverse (T) (resp. longitudinal (L)) modes correspond to linear superpositions of states with odd (resp. even) $N$ numbers of flipped spins (in red) between two spinons (dashed  purple lines), yielding a total spin $S_z=\pm 1$ (resp. $S_z=0$). In  both cases, the lowest energy excitation consists mostly in the state  with the minimal $N$ value, with a smaller contribution for the higher  $N$ states in the linear superposition. Neutron scattering can detect  onl
y $\Delta S=0,\pm 1$ type of excitations, hence pairs of spinons,  which individually carry a spin 1/2. {\bf c}, Numerically calculated magnetic  excitations of \bacovo\ in zero field, which show the Zeeman ladders (details in the last section of the article). The parameters of the calculations determined from a comparison with the zero-field measured excitations \cite{grenier2015} are $J=5.8$ meV, $\epsilon=0.53$, and $J'=0.17$ meV and were used for all the subsequent calculations in a transverse field. The color scale is in arbitrary units.}
\label{fig1}
\end{center}
\end{figure}

The cobalt oxide \bacovo\ indeed offers the unique opportunity to study this physics. This material exhibits screw chains of Co$^{2+}$ rotating around the 4-fold ${\bf c}-$axis. The magnetic moments of the Co$^{2+}$, in a distorted octahedral environment, are described as highly anisotropic effective spins $S=1/2$ (see Fig.~\ref{fig1}a) \cite{He2005}. In \bacovo, the presence of a small coupling between the spin chains additionally leads to a long-range ordering below the critical temperature $T_{\rm N}=5.5$~K. The order consists in an intra-chain AF arrangement of the magnetic moments pointing along the Ising ${\bf c}-$axis \cite{He2005,kimura2008a,canevet2013}.

Moreover, because of the original crystallographic structure of \bacovo, the magnetization local easy-axes of the Co$^{2+}$ ions are actually tilted away from the chain ${\bf c}-$axis by $\approx 5^{\circ}$ and rotate by 90$^{\circ}$ when moving along the 4-fold axis (see Fig.~\ref{fig1}a). This leads to a fully anisotropic g-tensor, which produces additional effective fields perpendicular to an applied transverse field~\cite{kimura2013}. These effective fields are different wether the applied field is along ${\bf b}$ (or equivalently ${\bf a}$) or along ${\bf a\pm b}$. In all cases, effective fields are produced along ${\bf c}$ whereas only in the former case a staggered field is induced along ${\bf a}$. Importantly, the critical field marking the end of the N\'eel phase determined from macroscopic measurements is $\mu_0H_c\approx10$~T for $\bf H\parallel \bf b$ and $\mu_0H_c'\approx40$~T for $\bf H\parallel \bf{a+b}$, the latter being much closer to the expected value corresponding to the magnetization satu
ration ~\cite{kimura2006,kimura2013,niesen2013}. For $\bf H\parallel \bf{b}$ only a small ferromagnetic component is induced at the critical field. As we will confirm in this article, this indicates that the effective staggered field induced along ${\bf a}$ opens an unconventional intermediate phase above $\mu_0H_c$.


The adequate model to describe \bacovo\ along with this phenomenology is the following:
\begin{align}
{\cal H} &= J \sum_{n,\mu} [\epsilon \left( S_{n,\mu}^x S_{n+1,\mu}^x + S_{n,\mu}^y S_{n+1,\mu}^y \right)+S_{n,\mu}^z S_{n+1,\mu}^z] \nonumber\\
&-  \sum_{n,\mu} \tilde g\mu_B {\bf H}\cdot{\bf S}_{n,\mu}
 + J' \sum_n \sum_{\mu,\nu (\mu \ne \nu)} S_{n,\mu}^z S_{n,\nu}^z
\label{eq1}
\end{align}
The first term is the $XXZ$ Hamiltonian where ${\bf S}_{n,\mu}$ is a spin 1/2, $\mu$ and $n$ are chain and site indexes, $J>0$ is the antiferromagnetic (AF) intra-chain interaction, $\epsilon$ the anisotropy parameter ($0 < \epsilon<1$ in our case). The action of a magnetic field yields the second term with $\mu_B$ the Bohr magneton, $\tilde g$ the Land\'e tensor and ${\bf H}$ the external magnetic field. The last term arises from the weak inter-chain coupling $J'$. In the specific case of $\bf{H} \parallel b$, the case that we shall consider below, the total (external + effective) magnetic field at site $n$ writes:
\begin{equation}
{\tilde g} {\bf H}_n = H \left[g_{yx} (-1)^n ~{\bf x} + g_{yy} ~{\bf y} + g_{yz} \cos \left( \pi \frac{2n-1}{4} \right) ~{\bf z}\right]
\label{eq2}
\end{equation}
with ${\bf x}={\bf a}$, ${\bf y}={\bf b}$ and ${\bf z}={\bf c}$.


Spectroscopic studies in zero field showed~\cite{kimura2007,grenier2015} that \bacovo\ does not host a classical N\'eel state at low temperature, in the sense that it lacks conventional spin wave excitations. Instead, the excitations consist of 2-spinons bound states confined by the interchain coupling (see Fig.~\ref{fig1}c for the calculated spectrum), leading  to series of long-lived gapped discretized modes (called Zeeman ladders) strongly dispersing along the ${\bf c}-$axis (and only weakly along the ${\bf a}$ and ${\bf b}$ directions). 
As depicted in Fig.~\ref{fig1}b, these bound states exist in two different flavors, depending on the parity of the number $N$ of flipped spins between two domain walls: the modes with $N$ odd (resp. even) carry a spin $S_z=\pm1$ (resp. $S_z=0$)~\cite{ishimura1980}. In \bacovo, the quasiparticles are described as linear combinations of those spinon pairs that we label using the index $j$, in ascending order, as $|j,S_z=\pm1 \rangle $ and $|j,S_z=0\rangle $. In neutron scattering experiments, the $|j,S_z=\pm1 \rangle $ modes come as transverse (T) excitations (spin fluctuations in the plane perpendicular to the direction of the ordered magnetic moment, hence to the $\bf c-$axis in zero field), while the $|j,S_z= 0\rangle $ modes come as longitudinal (L) ones (spin fluctuations parallel to the ordered moment). Note that $\epsilon$ allows the walls to flip by two sites. As a result, the $S_z=\pm1$ and $S_z=0$ sectors, hence the T and L modes, are decoupled. This original excitation spectrum occurs in \bacovo\ due
 to the sizable interchain interactions and to moderate Ising anisotropy~\cite{grenier2015}. Finally, it is also observed in the related compound SrCo$_2$V$_2$O$_8$ \cite{wang2015,wang2016,bera2017}.

\begin{figure*}[htb]
\begin{minipage}[h]{\linewidth}
\begin{center}
\includegraphics[width=18 cm]{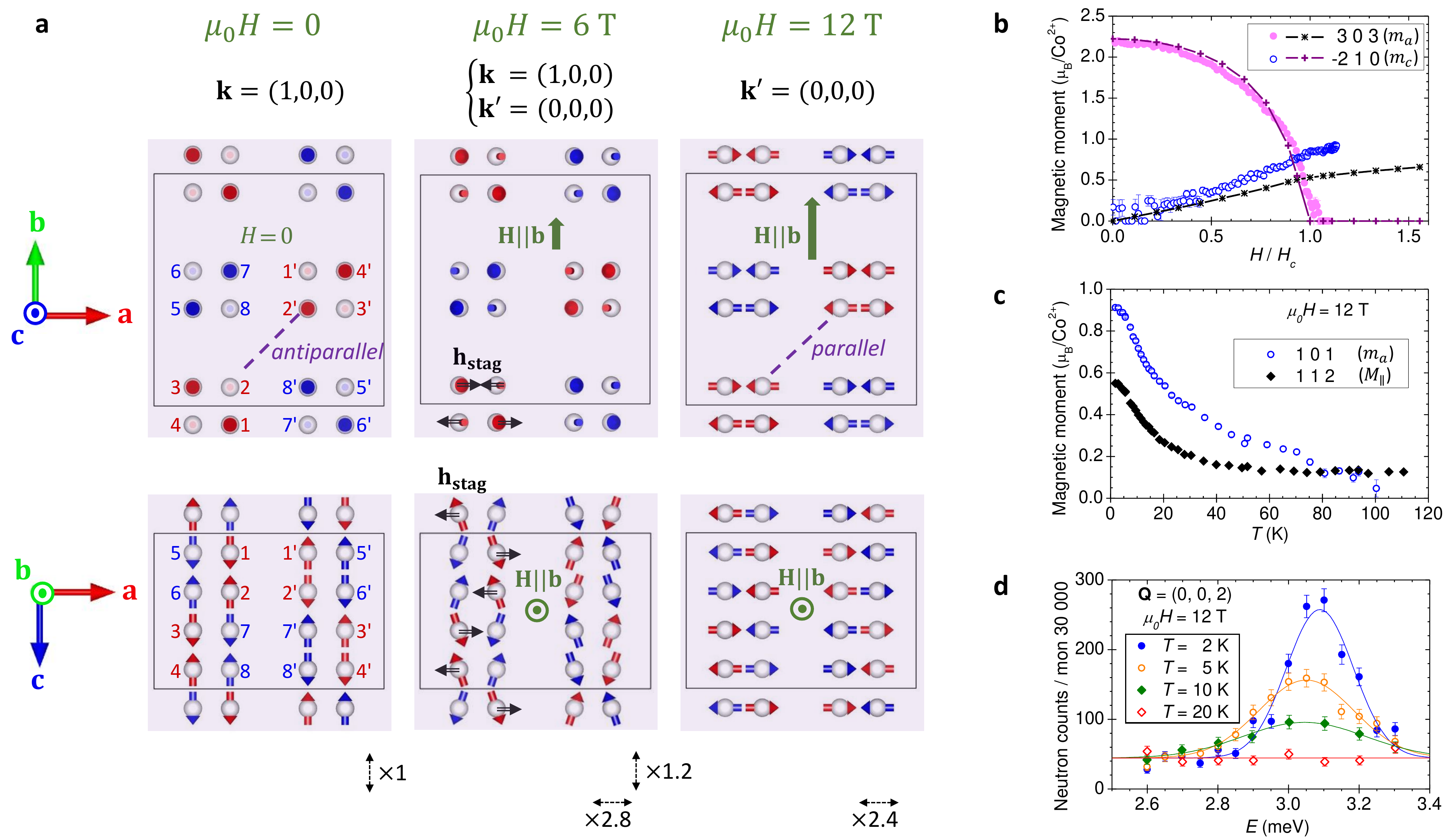}
\caption{{\bf Static properties in a transverse magnetic field.} {\bf a}, Magnetic structures of \bacovo\ in a uniform  field $\bf{H} \parallel b$ shown in the $({\bf a}, {\bf b})$ plane (top) and in the $({\bf a}, {\bf c})$ plane (bottom) at 0, 6, and 12~T, obtained from neutron diffraction experiments performed on the D23 diffractometer. The two kinds of chains are plotted in red and blue, respectively, and the Co atoms are numbered in the zero-field panels, like in Ref.~\cite{canevet2013}. For a better visualization, the amplitude of the magnetic moments was multiplied along the {\bf a}  and {\bf c} axes by the values indicated below the figures. The staggered field ${\bf h_{\rm stag}}$ is represented with black arrows, only in the panels corresponding to the 6 T magnetic structure for one chain. The end collinear magnetic structures have different propagation vectors, {\bf  k} and {\bf k'} indicated above the figures, and thus have different  interchain magnetic arrangements: For $H=0$, the bonds along ${
\bf  a}\pm{\bf b}$ are AF aligned while they are ferromagnetically (FM) aligned for  $\mu_0H=12$~T (see dashed purple line). Consequently, the contributions  to the total energy of bonds along ${\bf a}$ and ${\bf b}$ cancel each  other for $H=0$, half of them being frustrated, at variance with the  non frustrated 12~T structure in which the bonds along ${\bf a}$ and  ${\bf b}$ are both AF aligned. The intermediate magnetic structure, at  $\mu_0H=6$~T, is described by both propagation vectors {\bf k} and {\bf  k'}, and is thus non collinear. {\bf b}, Field dependence of the staggered magnetic moments of the two competing magnetic phases determined at 1.5~K from  the square root of the intensity of the $\bar 2\,0\,1$ and $3\,0\,3$ pure magnetic reflections for the low and high field phases  respectively (pink and blue circles). These experimental staggered moments $m_c$ and $m_a$ are compared to the ones calculated by numerical simulations (red and blue crosses connected by solid lines). {\bf c}, Temperature de
pendence at $\mu_0 H=12$~T of the AF ($m_a$) and FM ($M_\parallel$) components of the high field phase, obtained from the square root of the intensity of the $1\,0\,1$ and $1\,1\,2$ reflections, respectively. {\bf d}, Temperature dependence of the lowest energy magnetic excitation at a zone center position and at $\mu_0 H=12$~T, measured on the IN12 three-axis spectrometer. The solid lines are Gaussian fits.}
\label{fig2}
\end{center}
\end{minipage}
\end{figure*}

\vspace{1\baselineskip}
{\bf Static properties in a magnetic field}

We now examine, for the static magnetic properties of \bacovo, the effect of the effective fields, in particular of the staggered one along ${\bf a}$, created by applying a magnetic field along ${\bf b}$. The results of the neutron diffraction experiments provide insight into the ground state evolution. Fig.~\ref{fig2}a display sketches of the refined magnetic structures measured at 0, 6 (below $H_c$) and 12 T (above $H_c$). With increasing $H$, the magnetic moments remain staggered but progressively rotate in the $({\bf a},{\bf c})$ plane, from the Ising ${\bf c}-$axis to the ${\bf a}-$axis precisely at the transition (see supplementary information). The field-evolution of the staggered components $m_a(H)$ and $m_c(H)$ of the ordered magnetic moments along ${\bf a}$ and ${\bf c}$ respectively are shown in Fig.~\ref{fig2}b. The ${\bf a}$ component increases at the expense of the ${\bf c}$ component that eventually vanishes at the transition.
This evolution of the staggered moment orientation from the Ising ${\bf c}$ axis to the ${\bf a}$ axis originates from an energetic compromise, between on the one hand the intrachain exchange interaction, and on the other hand the Zeeman energy gain due to the effective transverse fields \cite{sato2004}.

It is worth noting that, in principle, the effective field along ${\bf c}$ also induces a magnetic component of 0.04~$\mu_B$ at $\mu_0H=12$~T, as deduced from the calculations. Its value, however, is extremely small and has consequently no relevant role in the phase transition. 


Note that this transition is not a standard spin reorientation with a global rotation of the spins. Indeed, the relative orientations of the magnetic moments between neighboring chains are different in the zero field and in the high field structure, due to a different symmetry of the interchain interactions and the staggered field, respectively.
As shown in Fig.~\ref{fig2}a (see also supplementary information), to accommodate this competition, the spins rotate clockwise for half of the chains, and anti-clockwise for the other half, yielding a non-collinear intermediate magnetic structure. This subtle modification points out the role of the staggered field, which forces a magnetic structure that competes with the interchain interactions.

The peculiar nature of the high field phase at $\mu_0 H=$ 12 T is further illustrated in Fig.~\ref{fig2}c. It shows the measured temperature dependence of the staggered order parameter ($m_a$) compared to the uniform component ($M_{\parallel}$) induced along $\bf b$ by the field. In contrast with the usual abrupt drop of the order parameter expected for a temperature becoming larger than the interactions between magnetic moments above a critical field (see for instance Fig.~6 of reference \cite{canevet2013}), $m_a$ decreases smoothly up to high temperature. $m_a$ is thus induced by the staggered magnetic field as $M_{\parallel}$ is induced by the uniform magnetic field. The intrachain interaction $J$ is still effective in the high field phase, and gives rise to well-defined excitations as shown in Fig.~\ref{fig2}d. Note that these modes disappear between 10 and 20 K, at a much lower temperature than the staggered magnetization. These excitations are actually quite unconventional as described in the following.

\begin{figure*}[htb]
\begin{minipage}[h]{\linewidth}
\begin{center}
\includegraphics[width=18 cm]{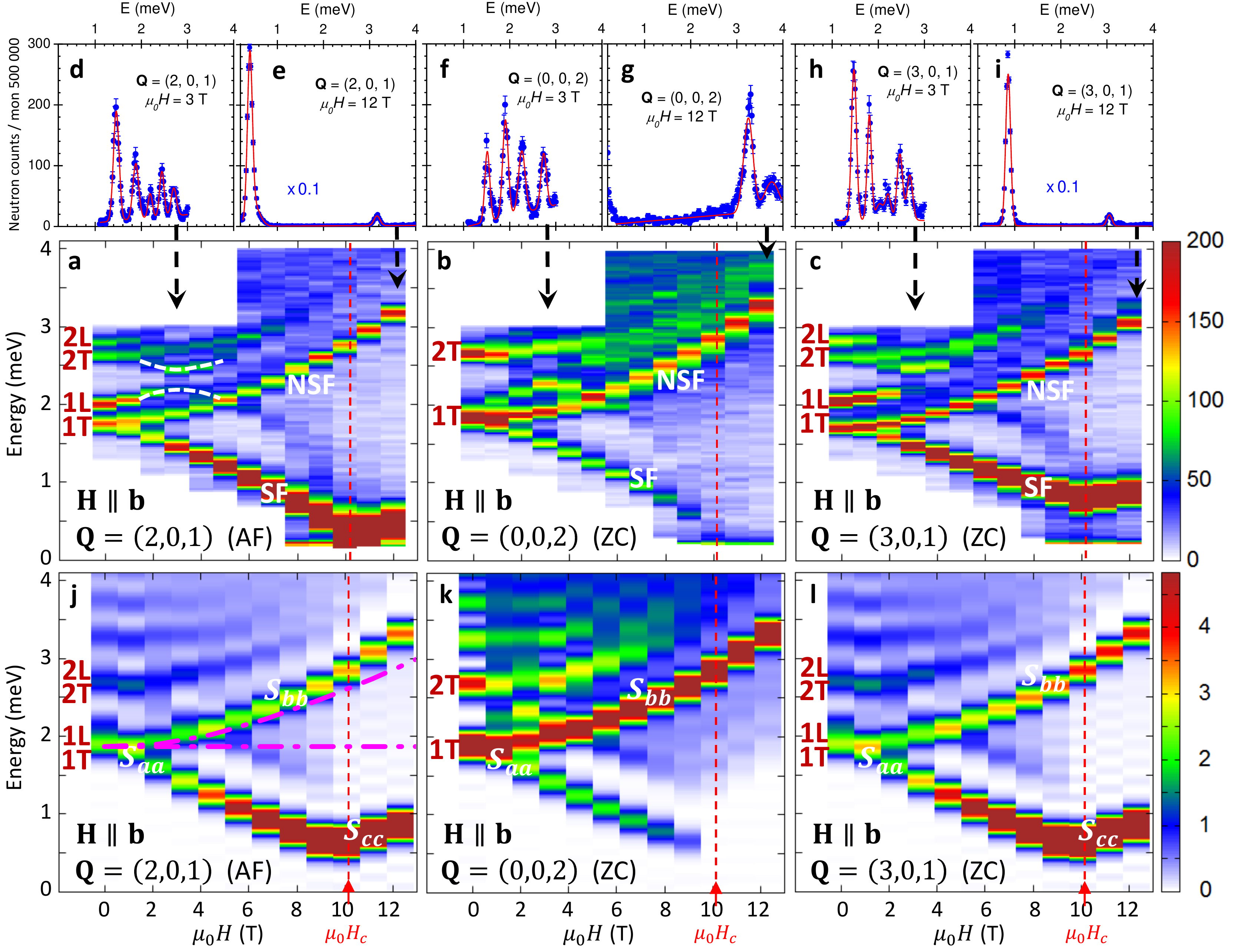}
\caption{{\bf Magnetic excitations in a transverse magnetic field.} {\bf a}-{\bf c}, Experimental intensity color maps showing the field dependence of the magnetic excitations in \bacovo\ for a transverse field. Three positions were investigated: AF ${\bf Q} = (2, 0, 1)$ and two ZC ${\bf Q} = (0, 0, 2)$ and ${\bf Q} = (3, 0, 1)$. The polarization (T or L) at zero field of the modes is indicated on the maps. Note that the color scale was truncated to about 10 to 15 times less than the maximum number of counts at high field for ${\bf Q} = (2, 0, 1)$ and $(3, 0, 1)$, for the weak modes to be visible. This gives the false impression that the lowest branch broadens as the field increases. The SF and NSF nature of the modes determined on IN12 using polarized neutrons is indicated. These three maps were obtained from energy scans performed on the ThALES triple-axis spectrometer at $T = 1.5$~K and every 1~T in the N\'eel phase ($0 \le \mu_0H \le \mu_0H_c \simeq 10$~T) and in the quantum phase above $H_c$ (critical fi
eld at the red dashed line). {\bf d}-{\bf i}, For each of these three scattering vectors, examples of such scans at $\mu_0H = 3$~T and $\mu_0H = 12$~T are shown above the corresponding map (blue points correspond to experimental data and red lines to fit by Gaussian functions). {\bf j}-{\bf l}, Theoretical intensity color maps to be compared to the experimental ones, obtained from iTEBD calculations of the $S({\bf Q},\omega)$ neutron scattering functions (see text). Here also, the color scale was truncated and both the critical field and the polarization (T or L) of the modes in zero-field are pointed out. In addition, the polarization (labeled $S_{aa}$, $S_{bb}$, and $S_{cc}$ for the $\bf a$, $\bf b$, and $\bf c$ directions, respectively) of the split branches arising from the 1T mode is indicated at $\mu_0H>0$: The polarization of the lowest branch evolves progressively from the $\bf a$ to the $\bf c$ direction and is thus always perpendicular to the ordered moment (see Fig.~\ref{fig2}a), hence transverse i
n the whole field range; the polarization of the higher branch is along $b$ (the field direction) in the whole field range, hence also transverse. The pink dashed lines in panel {\bf j} correspond to the field dependence these two branches would display in the absence of the staggered field along $\bf a$.}
\label{fig3}
\end{center}
\end{minipage}
\end{figure*}

\vspace{1\baselineskip}
{\bf Magnetic excitations in a magnetic field}

Figs~\ref{fig3}a-i show the measured evolution of the lowest modes ($|1,S_z=0,\pm1\rangle$ and $|2,S_z=0,\pm1\rangle$) of the Zeeman ladder, as a function of the transverse field ${\bf H} \parallel {\bf b}$, for several scattering vectors $\bf Q$: the AF point (2, 0, 1) and the two zone center (ZC) positions (0, 0, 2) and (3, 0, 1). By increasing $H$,  the zero-field $|1,S_z=\pm 1\rangle$ mode splits into two branches (see Figs~\ref{fig3}a-c). Note that the energy dependence of these two branches is not linear. The upper branch exhibits an upward variation up to $\mu_0H = 12$~T while the lower branch decreases down to $\mu_0H = 10$~T $= \mu_0 H_c$. At this field, this branch reaches its minimum energy before increasing again, as seen e.g. at the AF position ${\bf Q} =(2, 0, 1)$. The softening of the lower branch at $H_c$ thus marks the quantum phase transition, as already observed by Electron Spin Resonance (ESR)~\cite{okutani2015}. Note that a small energy gap of about $0.2$~meV is still present (see Fig.~\r
ef{fig3}a).  The width of these two modes remains resolution-limited, indicating that they still must be considered as long lived quasiparticles. The energy of the $|1,S_z=0\rangle$ mode is not constant with the field but increases with increasing field up to about $3$~T. At this field, an anti-crossing with the lowest branch of the upper $|2,S_z=\pm 1 \rangle$ mode occurs (see dashed white lines in Fig. \ref{fig3}a). As $H$ increases above $\mu_0H=3$~T, the lowest of the two hybridized branches broadens while its energy decreases, to finally disappear completely at the critical field.\\
\indent This field-dependence of the excitations is very different from the case of an external longitudinal field (parallel to the Ising axis) for which the $|j,S_z=\pm1\rangle$ and $|j,S_z=0\rangle$ excitations remain decoupled. In this case, the field produces a Zeeman splitting of the transverse excitations (linear field dependence) and has no effect on the longitudinal ones whose energy remains constant. This is indeed what is observed by ESR \cite{kimura2007} and inelastic neutron scattering in \bacovo\ (see supplementary information). The transverse field, on the other hand, allows the spinons to hop by one site and the $S_z=\pm 1$ and $S_z=0$ sectors are no more independent. As a result, the field creates a quantum overlap between the $|j,S_z=\pm1\rangle$ and $|j,S_z=0\rangle$ excitations (see Fig.~\ref{fig1}b). This hybridization process produces non-linearities to second order in $H$. In the present case, there are two kinds of transverse fields, the uniform one along {\bf b} and the staggered one a
long {\bf a}. The influence of the latter is the strongest one as shown in Fig.\ref{fig3}j since it produces the rapid decrease of the lower branch towards the critical field compared to almost no field dependence in its absence.\\
\indent A signature of the influence of the staggered field along {\bf a} is also visible in the field dependence of the intensity of the modes. The lowest energy mode displays a drastically different spectral weight evolution for the equivalent ZC $(0, 0, 2)$ and $(3, 0, 1)$ positions (see Figs~\ref{fig3}b,c). The latter gets more intense as the critical field is approached, while the former progressively vanishes. To understand this behavior, we performed inelastic neutron scattering measurements using polarized neutrons and polarization analysis in a vertical magnetic field parallel to the ${\bf b}$ axis on the IN12 triple-axis spectrometer. In this set-up, the non-spin-flip (NSF) and spin-flip (SF) scattering processes give information respectively about the spin fluctuations parallel to the field direction ${\bf b}$, and perpendicular to it (without discriminating between the ${\bf a}$ and ${\bf c}$ directions). The lowest branch of the split $|1,S_z=\pm1\rangle$ modes is found to be SF, hence polarized
within the $({\bf a},{\bf c})$ plane, while the upper branch is NSF, hence polarized along ${\bf b}$ (see supplementary information). Note that a geometrical factor enters the neutron cross section, reflecting the fact that only spin components perpendicular to the scattering vector ${\bf Q}$ contribute to the intensity. The decrease [resp. increase] of the spectral weight of the lowest energy mode for the (0, 0, 2) [resp. (3, 0, 1)] ZC can be explained by the change of polarization of the excitation, from parallel to $\bf a$ at low field to parallel to $\bf c$ at the critical field and above. This result can be understood by the rotation of the ordered moment from $\bf c$ to $\bf a$, as established by the diffraction results, the lowest excitation branch thus conserving its transverse character in the whole field range.

\vspace{1\baselineskip}

{\bf Topological nature of the transition and of the low energy excitations}

To determine the nature of the transition experimentally identified above, we performed numerical simulations of the $XXZ$ model in presence of an external magnetic field (see Eq.~\eqref{eq1}). The effects of the interchain interactions were taken into account by a mean field theory, in which an effective staggered field induced by the N\'eel order of the neighboring chains is determined self-consistently. We used an infinite time-evolving block decimation (iTEBD)~\cite{vidal2007} with the infinite boundary condition~\cite{phien2012} (see the supplementary information for details). Using the parameters $J=5.8$~meV, $\epsilon=0.53$ and $J'=0.17$ meV, close to those reported in the literature~\cite{kimura2013,grenier2015}, the results show excellent agreement with the experimental data and thus validate the model.
In zero magnetic field (Fig.~\ref{fig1}c), as discussed above, we reproduce the Zeeman ladders corresponding to the bound spinons \cite{grenier2015}. With the magnetic field, the results are shown in Fig.~\ref{fig2}b for the order parameter and in Figs~\ref{fig3}j-l for the excitation spectrum. In these calculations, we used the ratios $g_{yx}/g_{yy} \approx 0.40$ and $g_{yz}/g_{yy} \approx 0.14$ determined by Kimura {\it et al.}~\cite{kimura2013}, along with $g_{yy} \approx 2.35$, a value slightly different from the value of 2.75 determined in Ref.~\cite{kimura2013}, in order to agree with the measured critical field. The numerics describe extremely well the field dependence of the (staggered) magnetization along the chains. For the component perpendicular to the chains, the overall trend of the data is correctly given by the numerics but a global scaling factor seems to exist with the experimental data. The reason for this discrepancy could be due to factors such as: i) effect of temperature; ii) bigger sen
sitivity of this quantity on small uncertainties in the parameters, iii) treatment of the interchain interaction in the mean field theory. 
On the other hand, the results for the excitation spectrum (Figs~\ref{fig3}j-l) show a very good agreement with the data (Figs~\ref{fig3}a-c) accounting for the main modes observed experimentally. The numerics further validate the polarization of the modes and in particular the transverse nature of the lowest energy one. The rapid energy lowering of the lowest $|1,S_z=\pm1\rangle$ branch when the field is applied along ${\bf b}$ is confirmed numerically to be a consequence of the additional effective staggered field along $\bf a$ due to non diagonal components of the g-tensor~\cite{kimura2006,kimura2013}.

\begin{figure}
\begin{center}
\includegraphics[width=8.5 cm]{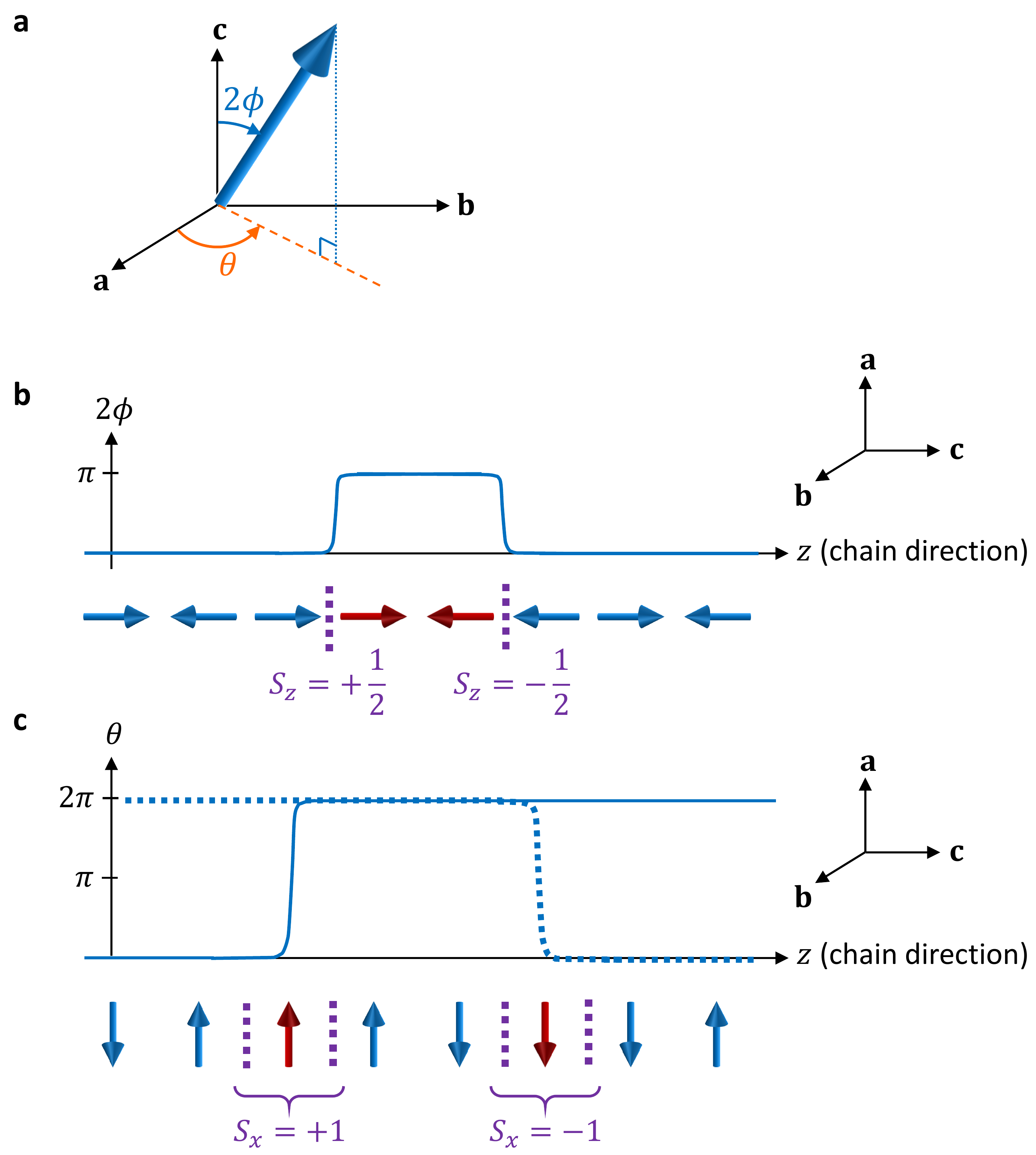}
\caption{{\bf Two dual topological objects.} {\bf a}, A qualitative interpretation of the two fields $\phi(z)$ and $\theta(z)$ entering in the field theory description. The quantum nature of a spin $1/2$ makes it impossible to determine both angles with an infinite accuracy, hence the fact that the two angles play a role similar to canonically conjugate variables. {\bf b}, Topological excitations in the low field phase. These excitations correspond to solitons in the field $4\phi(z)$ linking one of the minima of the $\cos 4\phi(z)$ to the next. Using the bosonization representation of the spin (see supplementary information), they can be identified with the spinon excitations (see also Fig.~\ref{fig1}b). They carry a spin $S_z = \pm 1/2$ corresponding to the topological index of the excitation. {\bf c}, In the high field phase, the elementary excitations now correspond to the solitons of the $\cos \theta(z)$, and are thus dual of the low field phase excitations. They carry an index of $S_x = \pm 1$.}
\label{fig4}
\end{center}
\end{figure}

The validation of the model of Eq.~(\ref{eq1}) from the numerics allows us to use field theory to describe the transition qualitatively, thereby determining its nature in a more transparent way.
Using the bosonization technique \cite{giamarchibook}, we obtain a dual-field double sine-Gordon model describing
\bacovo\ in an external field along ${\bf b}$:
\begin{align} \label{dsg}
 &\mathcal{H}^{\rm eff}=
   \frac{v}{2\pi}\int dz\Big[
   \frac{1}{K}\Big(\frac{d\phi(z)}{dz}\Big)^{2}
   +K\Big(\frac{d\theta(z)}{dz}\Big)^{2}\Big]\nonumber\\
   &-\frac{2\lambda}{(2\pi\alpha)^{2}}\int dz\cos4\phi(z)
   -\frac{g_{yx}\mu_{B}H}{\sqrt{2\pi\alpha}}\int dz\cos\theta(z),
\end{align}
(see supplementary information) where $v$ is the spinon velocity, $K$ the Luttinger parameter, $\lambda$ a contant having a dimension of energy and $\alpha$ a dimensionless constant~\cite{giamarchibook}. These parameters ($v,K,\lambda,\alpha$) are a function of $\epsilon$ and $H$, but there is no analytical representation. The effect of the Zeeman coupling with the uniform field along the ${\bf b}$ axis is renormalized into these parameters~\cite{AffleckOshikawa}. Since the Zeeman term of the four-site periodic field along the ${\bf c}$-axis is irrelevant, it does not appear in Eq.~(\ref{dsg}). The effect of this field is actually negligibly small.
$\phi(z)$ and $\theta(z)$ are dual bosonic fields that can be qualitatively identified with the polar and azimuthal angles of a staggered magnetization vector (see Fig.~\ref{fig4}a). In Eq.~(\ref{dsg}) the potential terms $\cos4\phi(z)$ and $\cos\theta(z)$ compete with each other and pin the fields $\phi(z)$ (resp. $\theta(z)$) for the low (resp. high) field phase. The expectation values $\langle\cos2\phi(z)\rangle$ and $\langle\cos\theta(z)\rangle$ correspond to the staggered magnetization along the ${\bf c}$ and ${\bf a}$ axes, respectively.

Excitations in a given phase correspond to the soliton of a pinned field (tunneling from one mininum of the cosine to the next), and carry a topological index (see Fig.~\ref{fig4}). In the low-field phase, $4\phi(z)$ is fixed to $2p\pi$ ($p$ is an integer). Thus, a low-energy excitation corresponds to the creation of a soliton-antisoliton pair in which $\phi(z)$ changes from $0$ to $2\pi/4$ (for the soliton) and from $2\pi/4$  back to $0$ (for the antisoliton) (see Fig.~\ref{fig4}b). The soliton and antisoliton are domain walls of the N\'eel order,  with spins \emph{reversed} with respect to the ground state between them. The solitons themselves can be identified with the spinons (see Fig.~\ref{fig1}b). The soliton-antisoliton would be deconfined in a single chain while confined in \bacovo\ due to the linear potential produced by interchain interaction. In the high-field phase the $\cos\theta(z)$ term dominates over $\cos4\phi(z)$ and fixes the $\theta(z)$ field.
The corresponding  soliton carries a spin $S_x=1$ (instead of $1/2$) since $\theta(z)$ changes from $0$ to $2\pi$ (instead of $0$ to $\pi$ for the field $2\phi$) (see Fig.~\ref{fig4}c). Note that important differences between the low and the high field phases exist. In the low field phase, the Hamiltonian contains $\cos(4\phi)$ while a physical observable such as the staggered part of $S^z$ depends on $\cos(2\phi)$. Thus the two parts of the soliton $2\phi=0$ and $2\phi=\pi$ can be distinguished by \emph{local} measurements of $S^z$, such as the string of overturned spins separating the two objects. However, in the high field phase, both the Hamiltonian and the physical observable $S^x$ depend on $\cos\theta$, making both sides of the solution identical far from the soliton. In terms of a \emph{local} measurement of $S^x$, the corresponding excitations would thus be local and would not carry a topological index. There is however a true topological order present since $\theta$ orders.  It could be detected thr
ough a quantity such as  $\langle e^{i \theta(z)/2} \rangle$, but would require nonlocal measurements as recently performed in cold atom systems~\cite{berg_string,endres_string_cold}. How to conduct such measurements for quantum spin systems in solid state is a challenging question.

The analysis of the properties of the modes in the experimental data and the agreement with numerics confirm that the quantum transition of the dual field double sine-Gordon model is indeed what is observed in \bacovo, providing an explanation of the rather mysterious field-induced transition and enlightening its topological nature. From a theoretical point of view, the study of the
transition itself is a challenging problem. The nature of the transition depends on the precise periodicity of the cosines~\cite{BKT40years}. For the purely 1D Hamiltonian (3) special solvable points suggest an Ising transition~\cite{Tsvelik2012}, as also confirmed by a numerical calculation of the central charge. A complete study, in particular taking into account the effective 3D coupling beyond mean-field is still lacking.
\bacovo\ thus provides a remarkable experimental system in which this transition can be tuned and studied in a controlled way.

More generally, quantum spin systems have been a steady reservoir of experimental realizations of topological phases and transitions, with in particular several realizations of the sine-Gordon model, or of exotic phases such as Tomonaga-Luttinger liquids. Our analysis of the transition in a uniform transverse magnetic field in \bacovo\, shows that they are also able to provide excellent and controlled realizations of more complex and yet challenging models from a theoretical point of view, confirming -- in addition to the own intrinsic interest of quantum magnets -- their place as quantum simulators of quantum correlated systems.\\


\noindent {\bf Acknowledgments}\\
We thank R. Ballou, C. Berthier, M. Horvati\'c, M. Klanj\v{s}ek, and S. Niesen for fruitful discussions, P. Courtois and R. Silvestre for their help in the sample co-alignment done at ILL prior to the experiment at PSI, E. Villard, B. Vettard, and M. Bartkowiak for their technical support during the Inelastic Neutron Scattering experiments on ThALES (ILL), IN12 (ILL) and TASP (PSI), respectively, J. Debray, A. Hadj-Azzem, and J. Balay for their contribution to the crystal growth, cut, and orientation. We acknowledge ILL and PSI for allocating neutron beam time. This work was partly supported by the French ANR Project DYMAGE (ANR-13-BS04-0013). ST is supported by the Swiss National Science Foundation under Division II and ImPact project (No. 2015-PM12-05-01) from the Japan Science and Technology Agency. MM acknowledges funding from the Swedish Research Council (VR) through a neutron project grant (Dnr. 2016-06955). TL acknowledges support by the Deutsche Forschungsgemeinschaft through CRC 1238 Project A02.

\section*{Author contributions}
\vspace{-0.2 cm}
All authors contributed significantly to this work. In details, sample preparation by PL, neutron scattering experiments and analysis by QF, BG, SP, and VS with the support of SR, LPR, MB, JSW, MM, and ChR, calculations by ST, SCF, and TG, Physical discussions with ChR, BC, and TL; Manuscript written by VS, SP, BG, QF, TG, and ST with constant feedback from the other co-authors.

\section*{Author information}
\vspace{-0.2 cm}
The authors declare no competing financial interests.
\\

\noindent {\bf Methods}\\

\noindent {\it Sample preparation \& experimental set up}

A \bacovo\ single-crystal was grown at Institut N\'{e}el by the floating zone method.~\cite{lejay2011} A 5~cm long cylindrical crystal rod, of about 4~mm diameter, was obtained by imposing the growth axis to be along the ${\bf b}$ crystallographic axis. One crystal piece of 10 mm long was cut from the rod for the diffraction experiment, while two crystal pieces of about 18 mm long were cut for the inelastic neutron scattering (INS) experiments.

The diffraction experiment was performed on CEA-CRG D23 single-crystal two-axis diffractometer with a lifting arm detector at Institut Laue Langevin (ILL). The sample, previously aligned with the ${\bf b}$ axis vertical on the Laue diffractometer OrientExpress at ILL, was installed on D23 in the CEA 12~T vertical field cryomagnet. A maximum transverse magnetic field of 12~T (${\bf H} \parallel {\bf b}$ and thus ${\bf H} \perp {\bf c}$ Ising axis) could be reached with a base temperature of 1.5~K. An incident wavelength of 1.28~\AA~was used, from a copper monochromator, thus allowing to measure $h\,0\,l$ and $h\,1\,l$ Bragg peaks with a maximum value of 17 for $h$ and 11 for $l$.

The INS experiments under a transverse magnetic field were performed on two cold-neutron triple-axis spectrometers, ThALES and FZJ-CRG IN12 at ILL. On ThALES,\cite{boehm2015} a PG(002) monochromator (resp. analyzer) was used to select (resp. analyze) the initial (resp. final) wave vector of the unpolarized neutron beam. On IN12, we used polarized neutrons, from a cavity transmission polarizer located far upstream in the guide, with an initial wave vector selected by a PG(002) monochromator, and polarization analysis, from a heusler analyzer (see Ref.~\onlinecite{Schmalzl2016} for a more detailed description of the standard polarized neutron setup on IN12). On both spectrometers, the energy resolution was of the order of 0.15 meV and the higher order contamination was suppressed by a velocity selector. The same cryomagnet as on D23 was used on both instruments, thus providing a maximum transverse field of 12~T at a base temperature of 1.5~K. Due to the high applied vertical magnetic field (up to 12~T), the ver
tical current of the Mezei spin flipper, placed just before the monochromator on IN12, was calibrated for every used value of the incident wave vector and of the magnetic field. The horizontal current was checked to be non sensitive to the applied field. The flipping ratios were ranging between 12 and 23, depending on the incident wave vector and magnetic field values. One of the two 200~mm$^3$ crystal pieces was used in both experiments and previously aligned with the ${\bf b}$ axis vertical on the triple-axis spectrometer IN3 at ILL, yielding a $({\bf a^\star}, {\bf c^\star)}$ horizontal scattering plane. Once the sample glued, the alignment was checked to be better than 1$^\circ$ on the neutron Laue diffractometer OrientExpress at ILL. All the INS data presented here were measured at a fixed final wave vector of 1.3~\AA$^{-1}$.


\clearpage
\widetext
\begin{center}
\textbf{\large Supplemental Information: \\Topological quantum phase transition \\ in the Ising-like antiferromagnetic spin chain BaCo$_2$V$_2$O$_8$}
\end{center}

\setcounter{equation}{0}
\setcounter{figure}{0}
\setcounter{table}{0}
\setcounter{page}{1}
\makeatletter
\renewcommand{\theequation}{S\arabic{equation}}
\renewcommand{\thefigure}{S\arabic{figure}}
\renewcommand{\bibnumfmt}[1]{[S#1]}
\renewcommand{\citenumfont}[1]{S#1}



\section{Single-crystal Neutron diffraction under a transverse magnetic field}

We report hereafter the details about the nuclear and magnetic structures refinement, based on the neutron diffraction measurements performed on the lifting-arm diffractometer CEA-CRG D23 at ILL, both in zero-field and under a transverse magnetic field applied along the $\bf b$ axis of the \bacovo\ single crystal up to 12 T. All data presented here were collected during the same experiment, and thus using exactly the same set-up (described in the method section of the main paper).\\

\bacovo\ crystallizes in the body-centered tetragonal $I4_1/acd$ space group with the following lattice parameters: $a = b = 12.444$~\AA\ and $c = 8.415$~\AA\ \cite{Wichmann1986}.\\

\begin{figure}[htb]
\begin{center}
\includegraphics[width=12.cm]{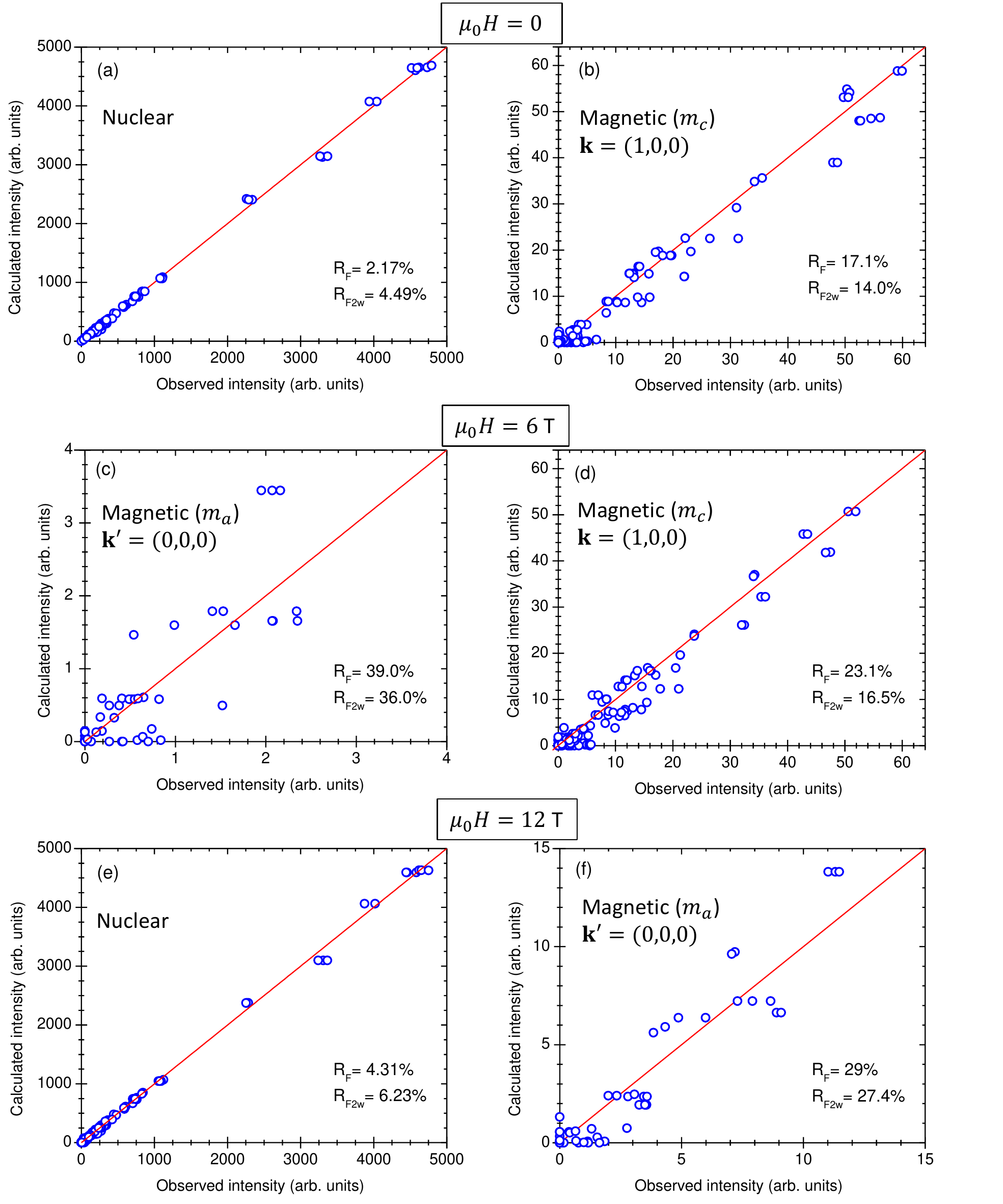}
\caption{Single-crystal diffraction data at 1.5 K, presented as calculated versus observed reflections intensities: (a) Nuclear structure refinement in zero field. (b) Magnetic structure refinement in zero field with the moments oriented along $\bf c$ [${\bf k} = (1,0,0)$]. (c) and (d) Magnetic structure refinement at 6 T for the component of the moments along $\bf a$ [${\bf k'} = (0,0,0)$] and along $\bf c$ [${\bf k} = (1,0,0)$], respectively. (e) Nuclear structure refinement at 12 T. (f) Magnetic structure refinement at 12 T with the magnetic moments oriented along $\bf a$ [${\bf k'} = (0,0,0)$]. The agreement factors are reported on the figures.}
\label{fig1SI}
\end{center}
\end{figure}

At $\mu_0H=0$, 288 nuclear reflections, allowed in the $I4_1/acd$ space group, reducing to 149 independent ones, were collected at $T=1.5$~K by performing rocking curves. The nuclear structure was then refined using the Fullprof software \cite{Rodriguez-Carvajal93} in order to determine the necessary information for the magnetic structure refinement (i.e. the scale factor, the $x$ coordinate and Debye-Waller factor of Co, the extinction parameters, and the $\lambda/2$ ratio). The calculated intensities $I_{calc}$ plotted in Fig.~\ref{fig1SI}a as a function of the observed ones $I_{obs}$ emphasize the quality of the fit. 103 magnetic reflections, associated to the ${\bf k}=(1,0,0)$ propagation vector and reducing to 48 independent ones, were then collected. Let us remind that they correspond to reflections $h\,k\,l$ with $h+k+l=2n+1$, that is, to Bragg positions for which the nuclear intensity is always null because of the body-centering of the crystallographic structure. The same magnetic structure as in Can\
'evet {\it et al.}\cite{canevet2013} was found, with a staggered moment $m_c=2.184(8)~\mu_B$/Co$^{2+}$ (see Fig.~\ref{fig1SI}b for the plot of $I_{calc}$ vs $I_{obs}$). Note that ${\bf k} = (1,0,0)$ implies an AF coupling in the diagonal ${\bf a} \pm {\bf b}$ direction, that is between two chains of the same nature (both described by a $4_1$ screw axis, plotted in red in Fig.~2a of the main paper, or by a $4_3$ screw axis, plotted in blue). Consequently, the magnetic structure presents an AF coupling along $\bf a$ and a FM one along $\bf b$ in the first domain (see top left panel of Fig.~2a in the main paper), while it is the reverse in the second domain.\\

The same nuclear reflections were then collected at $\mu_0H=12$\,T $>\mu_0H_c$, yielding the same crystallographic structure as in zero-field, to within the error bars, with comparable $R-$agreement factors. 48 magnetic reflections, with ${\bf k'} = (0,0,0)$, were then collected, reducing to 30 independent ones. This set of measured reflections (allowed by the lattice type) corresponds to those forbidden either by a screw axis or a glide plane of the $I4_1/acd$ space group. Nevertheless, some of them were not strictly null at zero-field because of the presence of a sizable $\lambda/2$ and/or mostly because of small defects in the crystal. As a result, they had to be collected in both phases and the difference between the $\mu_0H=12$~T collect and the $\mu_0H=0$ one was then used for the magnetic refinement. For this reason and because of the small magnetic signal, a counting rate of 30 seconds per point was used. The magnetic structure refinement was then performed, yielding a staggered magnetic moment $m_a=0
.91(2)~\mu_B$/Co$^{2+}$ (see Figs~\ref{fig1SI}e,f for the $I_{calc}$ vs $I_{obs}$ plots of the nuclear and magnetic refinements at 12~T). The propagation vector of the high field magnetic structure ${\bf k'} = (0,0,0)$ implies a FM coupling in the diagonal ${\bf a}\pm {\bf b}$ direction (see top right panel of Fig.~2a in the main paper). Consequently, an AF coupling both along ${\bf a}$ and along ${\bf b}$ now occurs, thus lifting the frustration and yielding a single magnetic domain.\\

The $\mu_0H=6$\,T $=0.6\,\mu_0H_c$ magnetic structure consists in a superposition of the ${\bf k} = (1,0,0)$ and ${\bf k'} = (0,0,0)$ phases, as shown by the field dependences plotted in Fig.~2b of the main paper. The same magnetic reflections as for $\mu_0H=0$ and $\mu_0H=12$~T were collected, with respective counting rates of 6 and 30 seconds. Here again, the difference with the zero-field phase was used for the second set of reflections. The results of the magnetic refinement is shown for both contributions (see Figs~\ref{fig1SI}c,d). The following values of the staggered components were found: $m_c=1.916(7)~\mu_B$/Co$^{2+}$ for the zero-field contribution and $m_a=0.44(2)~\mu_B$/Co$^{2+}$ for the high field one. The very small value of the latter component, in addition to the data treatment that had to be applied (difference with the zero-field data) explains the poorness of the fit. The non collinearity of the 6~T structure comes from the fact that it is a double $\bf k$ magnetic structure. It can be
simply understood by comparing the exchange couplings along $\bf a$ and $\bf b$ in the zero field structure (one is AF the other one is FM) to those in the high field phase (both are AF): As a result, half of the spins rotate clockwise and the other half anti-clockwise (see middle panels of Fig.~2a in the main paper).

\section{Neutron geometrical factor and Longitudinal Polarization Analysis}

As is well known, neutron scattering experiments probe the correlations between spin components perpendicular to the scattering vector ${\bf Q}$, denoted hereafter $\bf{S}_{\perp Q}$. As a result, the neutron scattering differential cross section reads:
\begin{equation}
\frac{d^2 \sigma}{d\Omega dE} \propto \langle \bf{S}_{\perp Q} \cdot \bf{S}_{\perp Q} \rangle
\end{equation}
where $\langle...\rangle$ is the thermal average.

\begin{figure}[htb]
\begin{center}
\includegraphics[width=10.3cm]{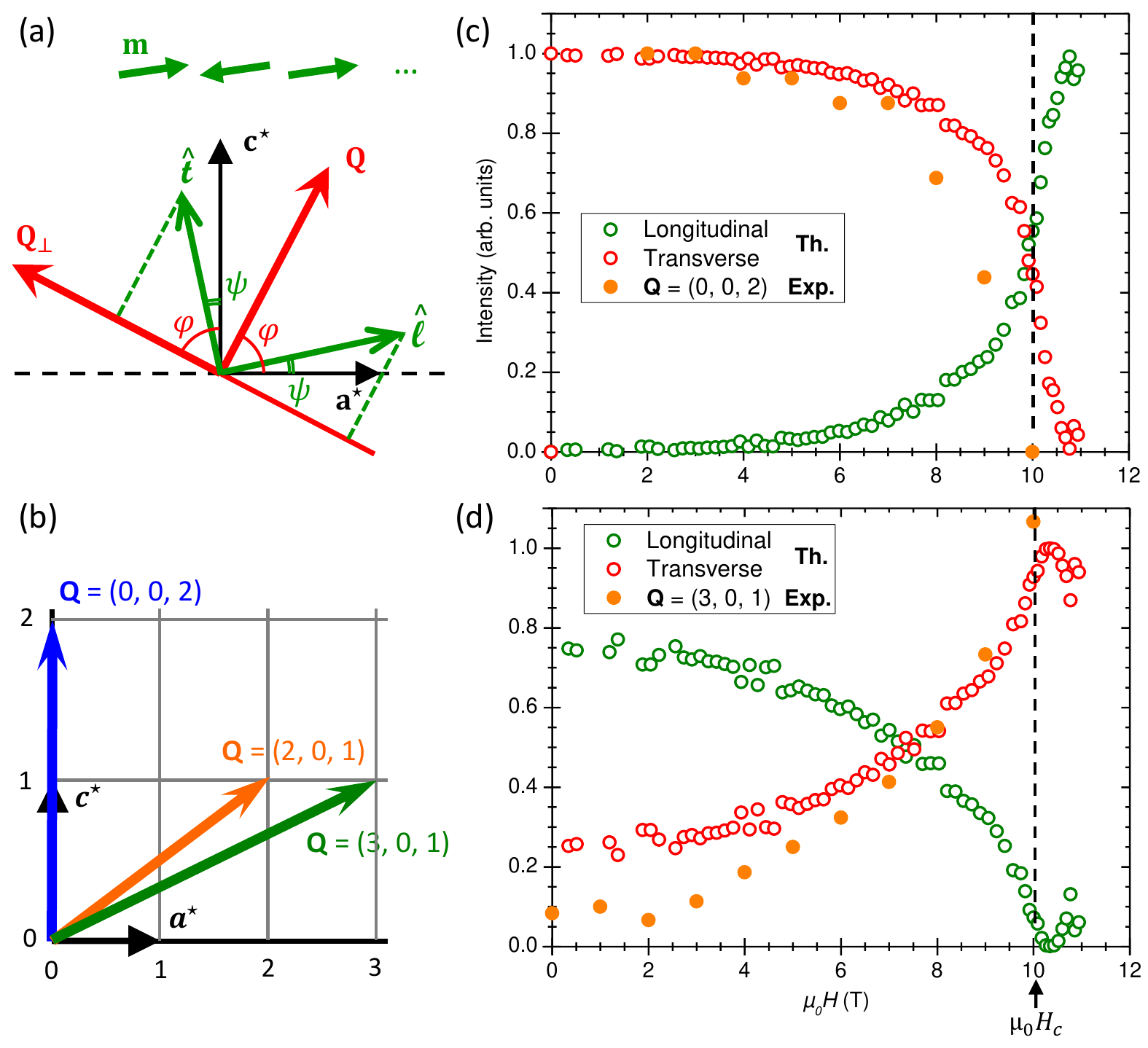}
\caption{(a) Rotating local frame $({\bf \hat \ell}, {\bf \hat b}, {\bf \hat t})$. The $\bf{\hat{\ell}}$ (resp. $\bf \hat{t}$) unitary vector is parallel (resp. perpendicular) to the ordered magnetic moment. The directions of the scattering vector $\bf Q$ and of $\bf{\hat{\ell}}$ with respect to ${\bf a^\star}$ are identified by the $\varphi$ and $\psi$ angles, respectively. (b) Location in the reciprocal lattice of the AF and ZC positions investigated. (c,\,d) Spectral weight of the lowest energy branch as a function of the transverse field for {\bf Q} = (0, 0, 2) and {\bf Q} = (3, 0, 1) ZC positions respectively, as measured (orange points), and calculated for longitudinal (green circles) and transverse (red circles) fluctuations. This clearly evidences the transverse nature of this excitation mode.}
\label{fig2SI}
\end{center}
\end{figure}

A convenient way to analyze the data is to consider longitudinal (L) and transverse (T) fluctuations with respect to the direction of the ordered moment ${\bf m}$ for a given field ${\bf H}$. Since the direction of ${\bf m}$ (in addition to its amplitude) changes with the field, a rotating frame $({\bf \hat \ell}, {\bf \hat b}, {\bf \hat t})$ is then introduced (see Fig.~\ref{fig2SI}a). $\hat{\bf{\ell}}$ is the (rotating) quantization axis, ${\bf \hat{t}}$ and ${\bf \hat{b}}$ are two orthogonal vectors such that ${\bf \hat{t}} = {\bf \hat{\ell}} \times {\bf \hat{b}}$. The vectors are defined as:
\begin{eqnarray*}
{\bf{\hat{\ell}}} &=& \frac{{\bf m}}{m} = \cos \psi~\frac{{\bf a^*}}{a^*} + \sin \psi~\frac{{\bf c^*}}{c^*} \\
{\bf \hat{t}} &=& -\sin \psi~\frac{{\bf a^*}}{a^*} + \cos \psi~\frac{{\bf c^*}}{c^*} \\
{\bf \hat{b}} &=& \frac{{\bf b^*}}{b^*}
\end{eqnarray*}
where $\psi$ is the $({\bf a^*},{\bf m})$ angle. Using this frame, the differential cross section reads:
\begin{equation}
\frac{d^2 \sigma}{d\Omega dE} \propto \sum_{x,y=t,b,\ell} \left < S_x \left(\delta_{xy}-\frac{Q_xQ_y}{Q^2}\right)
S_y \right >
\end{equation}

\begin{figure}[htb]
\begin{center}
\includegraphics[width=10cm]{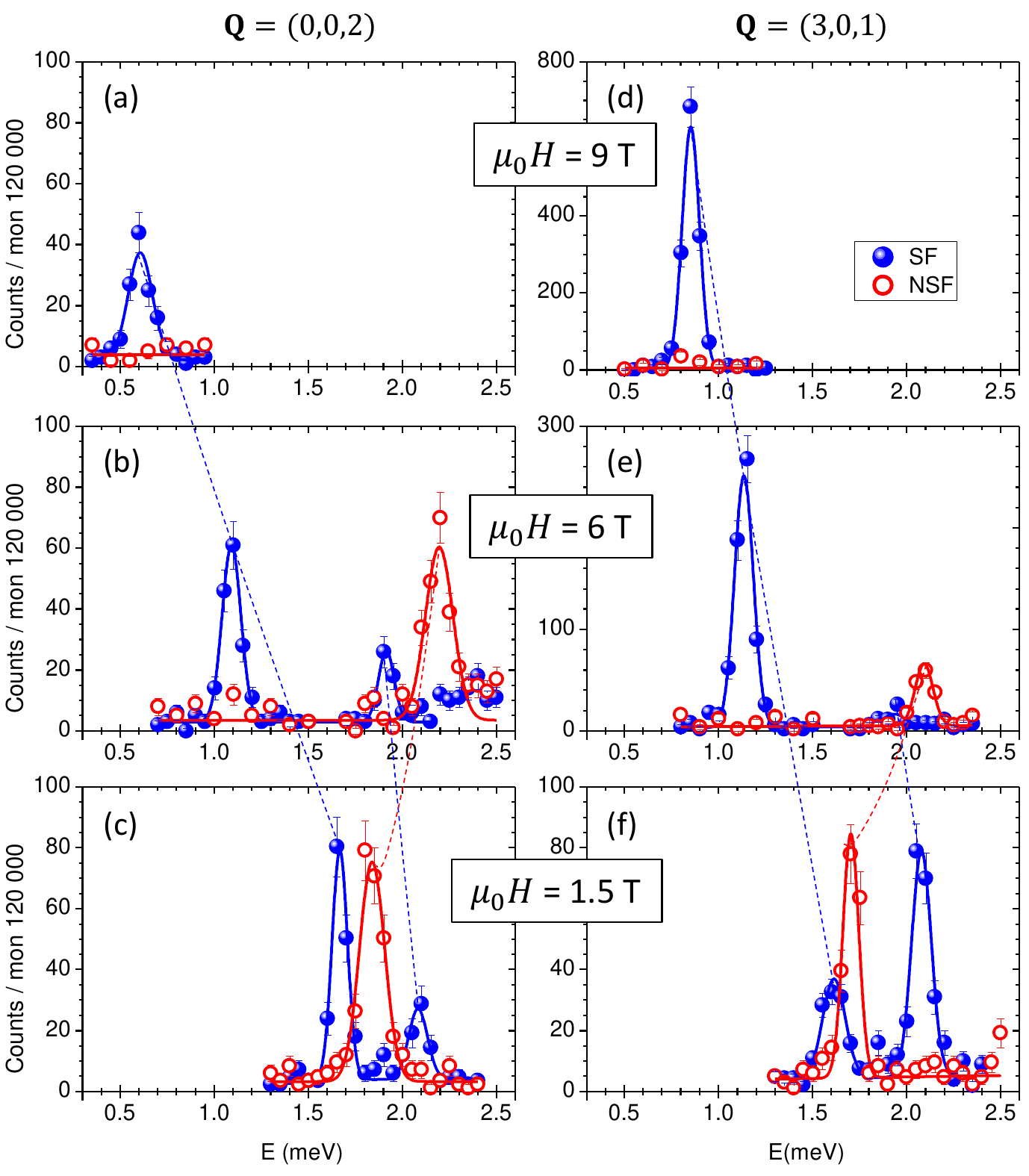}
\caption{Energy scans measured in \bacovo\ at T = 1.5 K on IN12 using polarized neutrons (open and closed symbols) fitted to Gaussian functions (solid lines), for (a-c) {\bf Q} = (0, 0, 2) and (d-f) {\bf Q} = (3, 0, 1), in the non spin-flip (NSF) and spin-flip (SF) channels. These data were obtained for three different values of an applied transverse magnetic field ${\bf H} \parallel {\bf b}$: $\mu_0H=1.5$~T (lower panels), $\mu_0H=6$~T (central panels), $\mu_0H=9$~T (upper panels). The dashed lines help to visualize the evolution of the three modes with the magnetic field, and in particular the variation in the intensity of the lowest mode decreasing with the field for {\bf Q} = (0, 0, 2) and increasing for {\bf Q} = (3, 0, 1). Note that an enlarged vertical scale (neutron counts) was used for panels (d) and (e), as compared to the four other panels.}
\label{fig3SI}
\end{center}
\end{figure}

For a ${\bf Q}$ vector making an angle $\varphi$ with ${\bf a^*}$ in the $({\bf a^*}, {\bf c^*})$ scattering plane, we have $Q_t=Q \sin{(\varphi-\psi)}, Q_b=0, Q_\ell=Q \cos{(\varphi-\psi)}$, hence:
\begin{eqnarray}
\frac{d^2 \sigma}{d\Omega dE} &\propto &\left[1-\sin^2{(\varphi-\psi)}\right] \langle S_t S_t \rangle + \langle S_b S_b \rangle + \left[1-\cos^2{(\varphi-\psi)}\right]\langle S_\ell S_\ell \rangle - 2 \cos{(\varphi-\psi)}\sin{(\varphi-\psi)} \langle S_t S_\ell \rangle \nonumber\\
&\propto & \cos^2{(\varphi-\psi)}  \langle S_t S_t \rangle + \langle S_b S_b \rangle + \sin^2{(\varphi-\psi)} \langle S_\ell S_\ell \rangle - 2 \cos{(\varphi-\psi)}\sin{(\varphi-\psi)} \langle S_t S_\ell \rangle
\end{eqnarray}
\\

The neutron cross section for longitudinal and transverse fluctuations acquires a ``geometrical factor'' $\sin^2 (\varphi-\psi)$ and $\cos^2 (\varphi-\psi)$ respectively. Note that the cross term $\langle S_t S_\ell \rangle$ is usually small and is thus neglected. A stringent comparison between the measured field dependence of the spectral weight for ${\bf Q}=(0,0,2)$ and $(3,0,1)$ (see Figs~3b,c of the main article) and the simulation using the above geometrical factors along with the diffraction data is shown in Figs~\ref{fig2SI}c,d. While the magnetic moments rotate from the ${\bf c}-$axis ($\psi=\pi/2$) to the ${\bf a}-$axis ($\psi=0$), the low energy branch originating from the split $|1,S_z=\pm1\rangle$ modes indeed follows the transverse geometrical factor with in particular $\varphi=\pi/2$ for ${\bf Q}=(0, 0, 2)$ and $\varphi\approx 26.3^{\circ}$ for ${\bf Q}=(3, 0, 1)$.\\

A spin polarized beam (as available on the triple-axis FZJ-CRG IN12 cold neutron spectrometer at ILL) provides additional information. In the present set-up (see Methods in the main paper for more details), the neutron beam is polarized by a cavity transmission polarizer with an initial wavevector selected by a PG(002) monochromator. The vertical magnetic field at the sample position then drives the spin of the incident neutrons parallel to the ${\bf b}-$axis. The scattered intensity is then analyzed (using a heusler analyzer), to separate the spin-flip (SF) and non-spin-flip (NSF) contributions, corresponding respectively to processes where the neutron spin has flipped or not (see Fig.~\ref{fig3SI}). It turns out that the NSF scattering cross section probes the correlations between spin fluctuations perpendicular to {\bf Q} and parallel to the incident neutrons polarization direction ({\bf b} in the present case), while the SF contribution probes the correlations between spin fluctuations perpendicular to bo
th {\bf Q} and the polarization direction. As a result, owing to the above framework, the NSF intensity will be directly proportional to $\langle S_b S_b \rangle$ while the SF intensity will probe $\cos^2{(\varphi-\psi)}  \langle S_t S_t \rangle + \sin^2{(\varphi-\psi)} \langle S_\ell S_\ell \rangle$. We conclude that at these positions, the upper branch of the $|1,S_z=\pm1\rangle$ Zeeman ladder mode is polarized along the ${\bf{\hat{b}}-}$axis while the lower one is polarized along the ${\bf{\hat{t}}-}$axis.

\section{Magnetic excitations in a longitudinal magnetic field}

\begin{figure} [htb]
\begin{center}
\includegraphics[width=7cm]{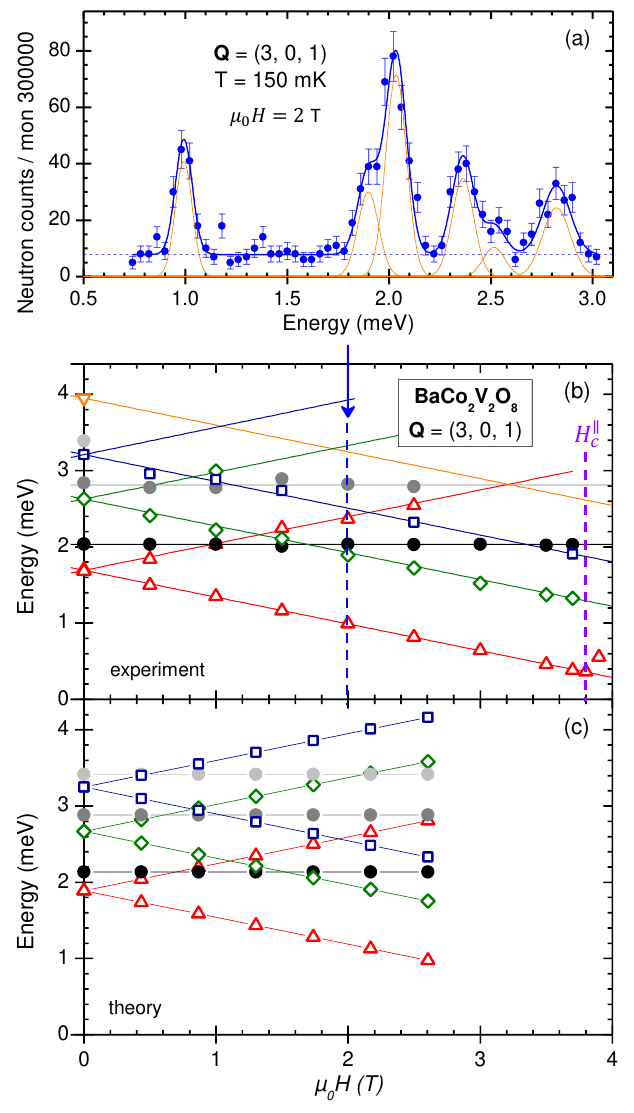}
\caption{(a) Example of ${\bf Q}-$constant energy scans performed on TASP in the N\'eel phase at $T=150$~mK on the ZC position ${\bf Q}=(3, 0, 1)$ for a longitudinal field $\mu_0H=2$~T (circles). The six observed modes were fitted to Gaussian functions (orange solid lines) on top of a constant background (blue dashed line), yielding the thick blue solid line for the total fit, in order to extract the energies of the various modes. Such scans were performed at various magnetic field values in the complete N\'eel phase ($0 \leq \mu_0H \leq 3.8$~T). (b) Resulting longitudinal field dependence of the energy positions of the excitations at ${\bf Q}=(3, 0, 1)$ (closed and open symbols for the L and T modes respectively). The field behavior of these modes is modeled by $E^{T,L}_{j\pm}=E^{T,L}_{j0} + g_{zz} S^{T,L}_z \mu_BH$ ($j=1,2,...$) with $g_{zz}=6.07$, $S^L_z=0$ (filled symbols), and $S^T_z=\pm 1$ (solid lines). The critical field is indicated by the dashed purple vertical line. (c) The calculations reproduce v
ery well the experimental field dependence of the excitations.}
\label{fig4SI}
\end{center}
\end{figure}

An inelastic neutron scattering experiment was performed on the cold-neutron triple-axis spectrometer TASP at Paul-Scherrer Institute (PSI) aiming at measuring the magnetic excitations of \bacovo\ in a longitudinal magnetic field ${\bf H}_c^\parallel$ (applied along the Ising ${\bf c}-$axis). Two crystal pieces were used: prior to the experiment, they were first individually aligned with the ${\bf b}$ axis vertical using X-ray Laue diffraction at Institut N\'eel, then co-aligned to within a rotation angle of 0.2$^\circ$ around the vertical direction, by using the hard X-ray Laue diffractometer of the neutron optics group at ILL. A 7~T horizontal field cryomagnet was installed on TASP, equipped with a dilution insert, allowing to reach a maximum field of 4.2~T only (due to strong forces exerted by the stray fields onto the sample table, still magnetic at that time) and a base temperature of 150~mK. The magnetic field was applied along the ${\bf c^\star}$ axis of the $({\bf a^\star}, {\bf c^\star})$ scattering
plane, thus corresponding to a longitudinal field. The data were measured at a fixed final wave vector of 1.2~\AA$^{-1}$ using graphite PG(002) monochromator and analyzer. An energy resolution of 0.075 meV was then achieved. A beryllium filter was installed after the sample to remove high order contaminations. No other ZC nor AF positions than the ${\bf Q}=(3, 0, 1)$ one were accessible in this set-up over a sufficiently large energy-range due to the four pillars building the horizontal cryomagnet (only four sectors of about 45$^\circ$ are accessible for the incident and diffracted beams).

The results presented here (see Figs~\ref{fig4SI}a,b) were obtained at 150 mK up to the critical field $H_c^\parallel$ at which the N\'eel phase transforms into an incommensurate longitudinal Spin Density Wave (SDW) ordered phase, describable in terms of a Tomonaga-Luttinger liquid \cite{canevet2013,kimura2008a,kimura2008b,klanjsek2015}. At $H=0$, discretized modes, corresponding to Zeeman ladders with transverse and longitudinal character, can be observed at the zone center (ZC) position ${\bf Q}=(3, 0, 1)$ as in our previous study~\cite{grenier2015}. Up to a critical field $\mu_0 H_c^\parallel\simeq3.8$~T at 150 mK, the $|j,S_z=0\rangle$ modes remain at the same energy position, whereas the $|j,S_z=\pm1\rangle$ modes get linearly split (see Fig.~\ref{fig4SI}b). This field-dependence is due to the fact that the longitudinal magnetic field keeps independent the $S_z=0$ and $S_z=\pm1$ sectors and thus does not mix the corresponding modes. The zero field transverse excitations simply experience a Zeeman splitti
ng. The value of $g_{zz}$ determined from the fit of the linear field-dependency of the transverse modes is 6.07, in good agreement with the value of 6.2 derived from magnetization measurements \cite{kimura2006}. This observation confirms our previous identification of the longitudinal and transverse modes in zero field \cite{grenier2015} and is well reproduced by the numerical calculations (see Fig.~\ref{fig4SI}c) described in the main text and in the following sections. The transition to the incommensurate longitudinal phase occurs at $H_c^\parallel$ when the lowest energy transverse mode condenses (note that it is not strictly zero yet).


\section{Description of the excitations under a transverse magnetic field}

To understand the evolution of the excitations throughout the transition occurring under a transverse field applied along the ${\bf b}-$axis, it is instructive to rewrite the Hamiltonian given by:
\begin{eqnarray}
{\cal{H}} &=& J \sum_{n,\mu} [\epsilon \left( S_{n,\mu}^x S_{n+1,\mu}^x + S_{n,\mu}^y S_{n+1,\mu}^y \right)+S_{n,\mu}^z S_{n+1,\mu}^z] \nonumber\\
&- & \sum_{n,\mu} {\tilde g}\mu_B {\bf H}\cdot{\bf S}_{n,\mu} + J' \sum_n \sum_{\mu,\nu (\mu \ne \nu)} S_{n,\mu}^z S_{n,\nu}^z
\label{eq1}
\end{eqnarray}
in the rotating frame $({\bf \hat \ell}, {\bf \hat b}, {\bf \hat t})$ shown in Fig.~\ref{fig2SI}a. New operators $(\sigma^{\ell}, \sigma^b, \sigma^t)$ are introduced, along with $\sigma^t=(\sigma^++\sigma^-)/2$, $\sigma^b=(\sigma^+-\sigma^-)/2i$, the quantization axis $\hat{\ell}$ pointing along the ordered magnetic moment. This yields, for the intrachain part of the Hamiltonian ($XXZ$ and Zeeman terms of Eq.~\ref{eq1}):
\begin{eqnarray}
{\cal H} &= &\sum_n K^{\pm} (\sigma^+_n \sigma^-_{n+1}+\sigma^-_n \sigma^+_{n+1}) \nonumber\\
& + & K^{\pm\pm} (\sigma^+_n \sigma^+_{n+1}+\sigma^-_n \sigma^-_{n+1}) \nonumber\\
& + & K^{\ell \pm} (\sigma^\ell_n \sigma^+_{n+1}+\sigma^\ell_n \sigma^-_{n+1}) + K^{\ell \ell} \sigma^\ell_n \sigma^\ell_{n+1} \nonumber\\
& - &\mu_{\rm B} H h^\ell \sigma^\ell_n - \mu_{\rm B} H (h^+ \sigma^+_n + h^- \sigma^-_n) \label{eq2}
\end{eqnarray}
with
\begin{eqnarray*}
K^{\ell \ell}  &=& J (\epsilon \cos^2 \psi + \sin^2 \psi )\\
h^{\ell}    &=& (-1)^n g_{xy} \cos \psi + g_{xz} \cos \left(\pi \frac{2n-1}{4} \right) \sin \psi\\
\end{eqnarray*}
\begin{eqnarray*}
K^{\ell \pm} &=& J(1-\epsilon) \sin \psi \cos \psi \\
h^{\pm}    &=& \left[-g_{xy} (-1)^n \sin \psi + g_{xz} \cos \left( \pi \frac{2n-1}{4} \right) \cos \psi \right]/2 \pm i g_{xx}/2\\
K^{\pm}    &=& J \frac{\epsilon(1+\sin^2 \psi) + \cos^2 \psi}{4} \hspace{5 cm}\\
K^{\pm \pm} &=& J \left(\frac{1-\epsilon}{4}\right)\cos^2 \psi \\
\end{eqnarray*}

While complicated at first glance, this form of the Hamiltonian proves meaningful to understand, from a physical point of view, how the spinon bound states $|j,S_\ell\rangle$ evolve upon the field. For instance, $K^{\ell \ell}$ renormalizes the energies of the spinons from $J$ to $\epsilon J$. $K^{\pm}$ allows each kink forming the bound state to hop (independently) by {\it two} sites. This term does not mix the $S_\ell=0,\pm1$ sectors and plays the role of kinetic energy. It evolves from $\epsilon J/2$ in zero field up to $(1+\epsilon)/4~J$ above $H_c$. The staggered field contribution developing with the external field, that enters $h^{\ell}$ (but also $h^{\pm}$), behaves as a confinement potential, in a similar way as the inter-chain interaction $J'$ does in zero field. The remaining terms are more complicated: $K^{\ell\pm}$ and $h^{\pm}$ move the spinons by {\it one} site and entangle the $S_\ell=\pm 1$ with the $S_\ell=0$ sectors. This coupling is responsible for the mixing of the $|1,S_\ell=0\rangle$ an
d $|2,S_\ell=\pm1\rangle$ states described in the main article.

\begin{figure} [h]
\begin{center}
\includegraphics[width=8.5cm]{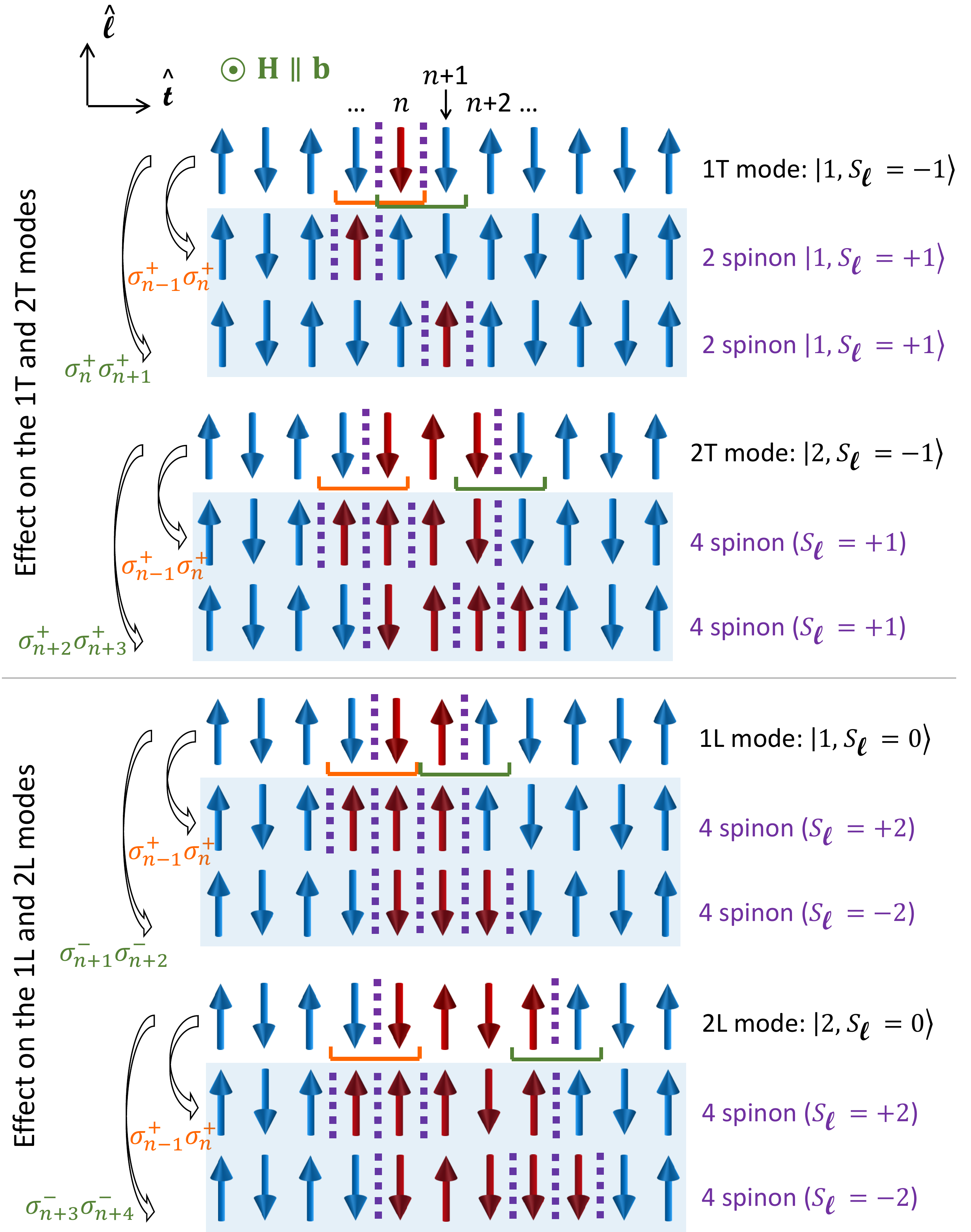}
\caption{Creation of multi-spinon states in the high field phase: cartoons explaining the effect of the $\sigma^\pm_n \sigma^\pm_{n+1}$ operator on the L and T excitations having the two smallest numbers $N$ of flipped spins, with the quantization axis $\bf{\hat{\ell}}$ along the staggered field direction ($\bf a$-axis).}
\label{fig5SI}
\end{center}
\end{figure}

The most peculiar term (absent in the isotropic Heisenberg case) is $K^{\pm\pm}$. It induces two spin-flips, changing $S_\ell$ by $\pm2$, and thus increases the number of spinons. Indeed, as shown in Fig.~\ref{fig5SI}, when acting on the $|j,S_\ell=0,\pm1\rangle$ states, this term essentially gives rise to 4 spinons states. This is likely the origin of the growing "incoherent" scattering that can be observed in Figs~3 and 4 of the main article. The cartoon shown in Fig.~\ref{fig5SI} shows further that $K^{\pm\pm}$ also couples the two $|1,S_\ell=\pm1\rangle$ states with each other, resulting in new eigenstates constructed as:
$$\frac{1}{\sqrt{2}}(|1,S_\ell=+1\rangle+|1,S_\ell=-1\rangle)$$ for the lower branch, and
$$\frac{1}{\sqrt{2}}(|1,S_\ell=+1\rangle+i|1,S_\ell=-1\rangle)$$ for the higher one.
Finally, $K^{\pm\pm}$ moves the two spinons by one site. Physically, this means that the new states get an extra kinetic energy provided the two spinons hop simultaneously. Although the physical mechanism is different, this quasiparticles become analogous to the kinetic bound state observed in the Ising-like FM compound CoNb$_2$O$_6$~\cite{coldea2010}.


\section{Numerical simulations}

\begin{figure}[h]
\includegraphics[width=0.8\textwidth]{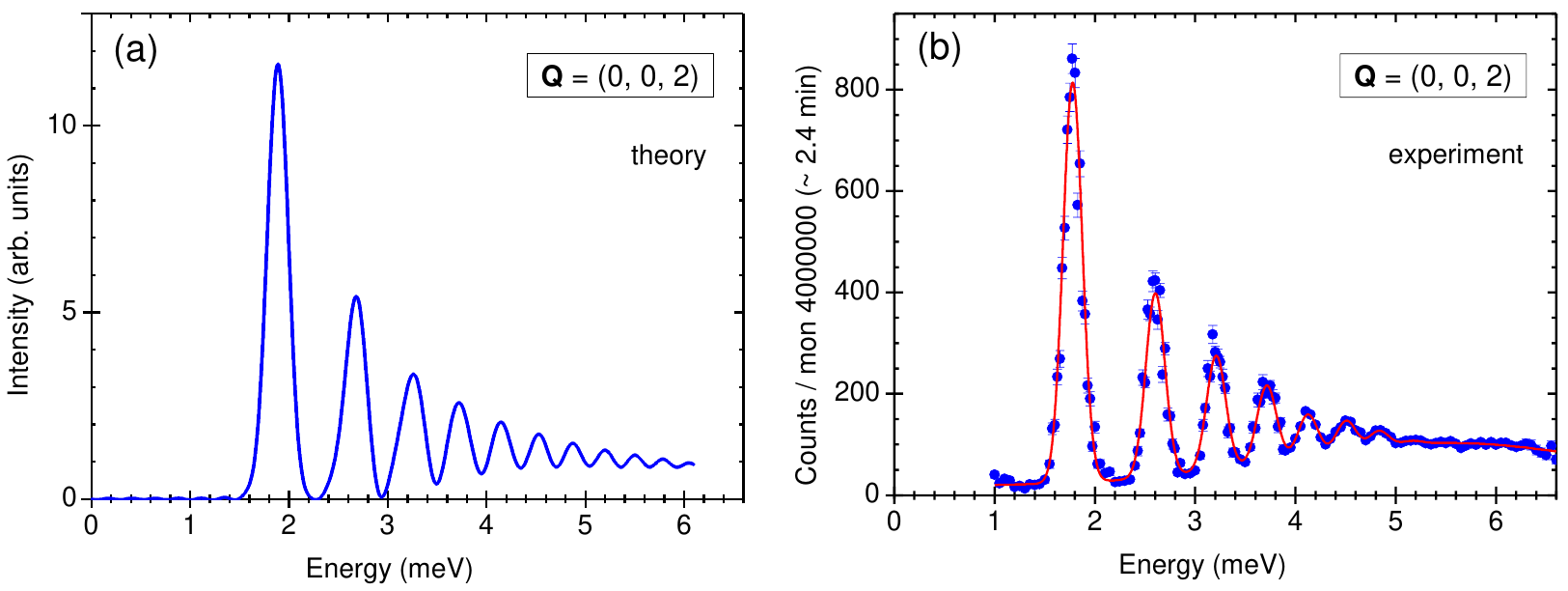}
\caption{(a) Intensity of scattered neutrons for ${\bf Q}=(0,0,2)$ at
 zero magnetic field calculated by iTEBD. The parameters are
$J=5.8\;\mathrm{meV}$, $\epsilon=0.53$, and
$g_{zz}\mu_{\mathrm{B}}h^{\mathrm{eff}}=0.061\;{\rm meV}$.
(b) Experimental results taken from Ref.~\cite{grenier2015} for comparison (solid symbols). The red line is a fit to the data explained in the corresponding article.  }
\label{fig6SI}
\end{figure}

In this section, we explain the method of numerical simulations for \bacovo.  This material consists of the stacking of Co chains, and each Co chain can be considered as a spin-1/2 Heisenberg model with Ising (easy-axis) anisotropy. The ground state of this Hamiltonian is a N\'eel ordered state. Here we take the contribution of interchain interaction $J'$ by a mean field approximation, which gives rise to an effective staggered magnetic field $\textbf{h}^{\rm eff}$ in the Hamiltonian (at $H=0$):
\begin{equation}
 {\cal H}=J\sum_{n}[\epsilon(S_{n}^{x}S_{n+1}^{x}
   +S_{n}^{y}S_{n+1}^{y})+S_{n}^{z}S_{n+1}^{z}]
   -g_{zz}\mu_{\mathrm{B}}h^{\rm eff}\sum_{n}(-1)^{n}S_{n}^{z}.
\end{equation}
The parameters $J,\epsilon$ and $h^{\rm eff}$ are determined in the way that the cross section for the scattering vector $\textbf{Q}=(0,0,2)$ (in the unit of $(2\pi/a,2\pi/b,2\pi/c)$, where $a,b,c$ are the lattice constants in the direction of the ${\bf a,b,c}$ axes) is reproduced. The differential neutron scattering cross section is represented as
\begin{equation}
  \frac{d^{2}\sigma}{d\Omega dE}\propto
   \frac{|\bf{q}'|}{|\bf{q}|}\sum_{\alpha,\beta=x,y,z}
   \Big(\delta_{\alpha\beta}-\frac{Q_{\alpha}Q_{\beta}}{|\bf{Q}|^{2}}\Big)
   |F(\textbf{Q})|^{2}\int dt\sum_{\textbf{r}}
   e^{i(\omega t-\textbf{Q}\cdot\textbf{r})}
   \langle S^{\alpha}(\textbf{r},t)S^{\beta}(\textbf{0},0)\rangle,
\label{eq:CrossSec}
\end{equation}
where $F(\bf{Q})$ is the magnetic form factor and $\bf{q},\bf{q}'$ are the initial and final wave vectors, respectively ($\bf{Q}=\bf{q}-\bf{q}'$). We calculate space-time correlation functions $\langle S^{\alpha}(\textbf{r},t)S^{\beta}(\textbf{0},0)\rangle$ using infinite time-evolving block decimation (iTEBD)~\cite{vidal2007} with the infinite boundary condition~\cite{phien2012}. In the calculations, the time is taken to be $0\leq t\leq 80J^{-1}$ with the discretization $dt=0.05J^{-1}$. The truncation dimension (i.e., dimension of matrix product states) is $\chi=60$. For the Fourier transform in Eq.~\eqref{eq:CrossSec}, the summation is taken over the actual positions $\bf{r}$ of Co atoms. The scattering cross section (i.e., intensity of scattered neutrons) for $\textbf{Q}=(0,0,2)$ at zero magnetic field calculated by iTEBD with the parameters $J=5.8\;\mathrm{meV}$, $\epsilon=0.53$, and $g_{zz}\mu_{\mathrm{B}}h^{\mathrm{eff}}=0.061\;\mathrm{meV}$ ($g_{zz}$ is the Land\'e's $g$ factor, and $\mu_{\rm B}=5.788\
times 10^{-5}\;\mathrm{eV/T}$ is the Bohr magneton) is shown in Fig.~\ref{fig6SI}. It reproduces very well the experimental result~\cite{grenier2015}. The effective staggered field $g_{zz}\mu_{\mathrm{B}}h^{\mathrm{eff}}$ originates from the interchain interaction. Using a mean field theory, we obtain $g_{zz}\mu_{\mathrm{B}}h^{\mathrm{eff}}=J'|\langle S_{j}^{z}\rangle|$, where $J'=0.17\;{\rm meV}$ and $|\langle S_{j}^{z}\rangle|=0.366$. The colormap of scattering intensity for $\textbf{Q}=(2,0,l)$ ($2\leq l\leq 3$) is also shown in the main article (Fig.~1c), which also agrees rather well with experiment (see Fig.~1 of reference \cite{grenier2015}) except for the anticrossing observed around 4-5 meV at $\textbf{Q}=(2, 0, 2.5)$.

When the uniform transverse field $H$ is applied, the system is subjected to effective fields in the $\bf x$ and $\bf z$ directions since the magnetic principal axes are inclined with respect to the crystal axes (see Fig.~1a in the main paper), which gives non-diagonal components to the g-tensor. With the external field parallel to the ${\bf b}-$axis, the Hamiltonian becomes, with $x=a$, $y=b$ and $z=c$:
\begin{align}
 {\cal H}&=J\sum_{n}[\epsilon(S_{n}^{x}S_{n+1}^{x}
   +S_{n}^{y}S_{n+1}^{y})+S_{n}^{z}S_{n+1}^{z}]\nonumber\\
   &-\mu_{\mathrm{B}}
     \sum_{n}\big(g_{zz}h^{\rm eff}(-1)^{n}S_{n}^{z}
     +(-1)^{n}g_{yx}HS_{n}^{x}+g_{yy}HS_{n}^{y}
     +\cos[(2n+1)\pi/4]g_{yz}HS_{n}^{z}\big).
\label{eq:HamilTransField}
\end{align}
Here we set the parameter $g_{yy}=2.35$, chosen slightly smaller than the value of Kimura {\it et al.}~\cite{kimura2013} in order to reproduce the critical field of the transition. The other parameters are $g_{yx}/g_{yy}=0.4$ and $g_{yz}/g_{yy}=0.14$ according to Ref.~\cite{kimura2013}. Since $h^{\mathrm{eff}}$ is introduced from the mean field approximation of the interchain interactions, $h^{\mathrm{eff}}$ is determined from the self-consistency equation $h^{\mathrm{eff}}(H)/h^{\mathrm{eff}}(0)=m_{z}(H)/m_{z}(0)$, where $m_{z}$ is the staggered magnetization of the N\'eel order along the ${\bf c}-$axis. Scattering cross sections calculated using Eq.~\eqref{eq:HamilTransField} under magnetic field are shown in the main article.

The chosen set of parameters is the best compromise to reproduce the staggered ordered moments along $\bf c$ and $\bf a$, the zero-field excitation spectrum, the value of the critical field and the field-dependence of the magnetic excitations.

\section{Dual field double sine-Gordon model}

The bosonization formula for spin operators is

\begin{align}
 S_{n}^{+}&=\frac{e^{-i\theta(z)}}{\sqrt{2\pi\alpha}}
   [(-1)^{n}+\cos2\phi(z)+\cdots]\\
 S_{n}^{z}&=\frac{1}{\pi}\frac{d\phi(z)}{dz}
   +\frac{(-1)^{n}}{\pi\alpha}\cos2\phi(z)+\cdots,
\end{align}

\noindent where $z=nc/4$ ($c$ is the lattice constant) and $\alpha$ is a nonuniversal constant~\cite{giamarchibook}. We take the lattice constant as unity $c=1$ hereafter.
The bosonized form of the $XXZ$ model without an external magnetic field is given by

\begin{equation}
 \mathcal{H}=\frac{v}{2\pi}\int dz\Big[
   \frac{1}{K}\Big(\frac{d\phi(z)}{dz}\Big)^{2}
   +K\Big(\frac{d\theta(z)}{dz}\Big)^{2}\Big]
   -\frac{2\lambda}{(2\pi\alpha)^{2}}\int dz\cos4\phi(z)
   +\cdots,
\end{equation}

\noindent where $v$ is the spinon velocity, $K$ is the Luttinger parameter and $\lambda$ is a contant having a dimension of energy~\cite{giamarchibook}. These parameters ($v,K,\lambda$) are a function of $\epsilon$, and they are renormalized when the uniform field along ${\bf b}$ is applied. However, there is no simple analytic form to represent $v,K,\lambda$ as a function of $\epsilon$ and the strength of the uniform field. Since the scaling dimension of the $\cos4\phi(z)$ term is $4K$ and $K\lesssim 1/2$ for $\epsilon<1$, the $\cos4\phi(z)$ term is relevant. Hence it opens an excitation gap in the system. The staggered field along $\bf a$ effectively induced by the application of transverse field along $\bf b$ also gives another relevant term

\begin{equation}
 \sum_{n}(-1)^{n}S_{n}^{x}=\frac{1}{\sqrt{2\pi\alpha}}\int dz\cos\theta(z)+\cdots,
\end{equation}

\noindent which has the scaling dimension $1/(4K)$. Thus the effective Hamiltonian in the bosonized field theory becomes

\begin{equation}
 \mathcal{H}^{\rm eff}=
   \frac{v}{2\pi}\int dz\Big[
   \frac{1}{K}\Big(\frac{d\phi(z)}{dz}\Big)^{2}
   +K\Big(\frac{d\theta(z)}{dz}\Big)^{2}\Big]
   -\frac{2\lambda}{(2\pi\alpha)^{2}}\int dz\cos4\phi(z)
   -\frac{g_{yx}\mu_{\rm B}H}{\sqrt{2\pi\alpha}}\int dz\cos\theta(z)
   +\cdots.
\label{eq:dsg}
\end{equation}

Since the Zeeman term of the four-site periodic field along the ${\bf c}$ axis is irrelevant, it does not appear in Eq.~(\ref{eq:dsg}). This model has two relevant cosine potentials in addition to the kinetic
term, which is then called a dual field double sine-Gordon model. For $\epsilon<1$ ($K\lesssim 1/2$), the $\cos\theta(z)$ term has a lower scaling dimension (i.e., is more relevant) than the $\cos4\phi(z)$ term. While the
$\cos4\phi(z)$ term is dominant for small $H$, the $\cos\theta(z)$ term dominates over the $\cos4\phi(z)$ term with increasing $H$ and the phase transition happens.


\begin{references}

\bibitem{Berezinskii}
Berezinskii, V. L.
Destruction of long-range order in one-dimensional and two-dimensional systems having a continuous symmetry group I. Classical systems.
{\it Sov. Phys. JETP} {\bf 32}, 493 (1971);
Destruction of long-range order in one-dimensional and two-dimensional systems having a continuous symmetry group II. Quantum systems.
{\it Sov. Phys. JETP} {\bf 34}, 610 (1972).

\bibitem{KosterlitzThouless1973}
Kosterlitz, J. M. \& Thouless, D. J.
Ordering, metastability and phase transitions in two-dimensional systems.
{\it Journal of Physics C: Solid State Physics} {\bf 6}, 1181 (1973).

\bibitem{BKT40years}
Jos\'e, J. V. 40 years of Berezinskii-Kosterlitz-Thouless theory. World Scientific (2013).

\bibitem{giamarchibook}
Giamarchi T. Quantum Physics in One Dimension
(Oxford University Press, Oxford, 2004).

\bibitem{Haldane1983}
Haldane, F. D. M.
Continuum dynamics of the 1-D Heisenberg antiferromagnet: Identification with the O(3) nonlinear sigma model.
{\it Phys. Lett. A} {\bf 93}, 464 (1983);
ibid.
Nonlinear Field Theory of Large-Spin Heisenberg Antiferromagnets: Semiclassically Quantized Solitons of the One-Dimensional Easy-Axis N\'eel State.
{\it Phys. Rev. Lett.} {\bf 50}, 1153 (1983).

\bibitem{Thouless1982}
Thouless, D. J., Kohmoto, M., Nightingale, M. P. \& den Nijs, M.
Quantized Hall Conductance in a Two-Dimensional Periodic Potential.
{\it Phys. Rev. Lett.} {\bf 49}, 405 (1982).

\bibitem{Hasan2010}
Hasan, M. Z. \& Kane, C. L.
Colloquium: Topological insulators.
{\it Rev. Mod. Phys.} {\bf 82}, 3045 (2010).

\bibitem{Haldane1988}
Haldane, F. D. M.
Model for a Quantum Hall Effect without Landau Levels: Condensed-Matter Realization of the ``Parity Anomaly''.
{\it Phys. Rev. Lett.} {\bf 61}, 2015 (1983).

\bibitem{Jotzu2014}
Jotzu, G., Messer, M., Desbuquois, R., Lebrat, M., Uehlinger, T., Greif, D. \& Esslinger T.
Experimental realization of the topological Haldane model with ultracold fermions.
{\it Nature} {\bf 515}, 237 (2014).

\bibitem{Kosterlitz1974}
Kosterlitz, J. M.
The critical properties of the two-dimensional $xy$ model.
{\it Journal of Physics C: Solid State Physics} {\bf 7}, 1046 (1974).

\bibitem{rajaraman_instanton}
R. Rajaraman.
Solitons and Instantons: An Introduction to solitons and Instantons in Quantum Field Theory (North Holland, 1982).

\bibitem{Jose1977}
Jos\'e, J. V., Kadanoff, L. P., Kirkpatrick, S. \& Nelson, D. R.
Renormalization, vortices, and symmetry-breaking perturbations in the two-dimensional planar model.
{\it Phys. Rev. B} {\bf 16}, 1217 (1977).

\bibitem{Fertig2002} Fertig, H. A. Deconfinement in the Two-Dimensional XY Model.
{\it Phys. Rev. Lett.} {\bf 89}, 035703 (2002).

\bibitem{GiamarchiSchulz1988} Giamarchi, T. \& Schulz, H. J.
Theory of spin-anisotropic electron-electron interactions in quasi-one dimensional metals.
{\it J. Phys. France} {\bf 49}, 819 (1988).

\bibitem{Lecheminant2002}
Lecheminant, P., Gogolin, A. O. \& Nersesyan, A. A.
Criticality in self-dual sine-Gordon models.
{\it Nucl. Phys. B} {\bf 639}, 502 (2002).

\bibitem{He2005} He, Z. Fu, D. Ky\^omen, T. Taniyama, T. \& Itoh, M.
Crystal Growth and Magnetic Properties of BaCo$_2$V$_2$O$_8$.
{\it Chem. Mater.} {\bf 17}, 2924 (2005).

\bibitem{kimura2008a} Kimura, S., Takeuchi, T., Okunishi, K., Hagiwara, M., He, Z., Kindo, K., Taniyama, T. \& Itoh, M.
Novel Ordering of an $S=1/2$ Quasi-1d Ising-Like Antiferromagnet in Magnetic Field.
{\it Phys. Rev. Lett.} {\bf 100}, 057202 (2008).

\bibitem{canevet2013} Can\'evet, E., Grenier, B., Klanj\v{s}ek, M., Berthier, C., Horvati\'{c}, M., Simonet, V. \& Lejay, P.
Field-induced magnetic behavior in quasi-one-dimensional Ising-like antiferromagnet BaCo$_2$V$_2$O$_8$: A single-crystal neutron diffraction study.
{\it P. Phys. Rev. B} {\bf 87}, 054408 (2013).

\bibitem{kimura2013} Kimura, S., Okunishi, K., Hagiwara, M., Kindo, K., He, Z., Taniyama, T., Itoh, M., Koyama, K. \& Watanabe, K.
Collapse of Magnetic Order of the Quasi One-Dimensional Ising-Like Antiferromagnet BaCo$_2$V$_2$O$_8$ in Transverse Fields.
{\it J. Phys. Soc. Japan} {\bf 82}, 033706 (2013).

\bibitem{niesen2013} Niesen, S. K., Kolland, G., Seher, M., Breunig, O., Valldor, M., Braden, M., Grenier, B. \& Lorenz, T.
Magnetic phase diagrams, domain switching, and quantum phase transition of the quasi-one-dimensional Ising-like antiferromagnet BaCo$_2$V$_2$O$_8$.
{\it Phys. Rev. B} {\bf 87}, 224413 (2013).

\bibitem{kimura2006} Kimura, S., Yashiro, H., Hagiwara, M., Okunishi, K., Kindo, K., He, Z., Taniyama, T. \& Itoh, M.
High field magnetism of the quasi one-dimensional anisotropic antiferromagnet BaCo$_2$V$_2$O$_8$.
{\it J. Phys.: Conf. Ser.} {\bf 51}, 99 (2006).

\bibitem{kimura2007} Kimura, S., Yashiro, H., Okunishi, K., Hagiwara, M., He, Z., Kindo, K. Taniyama, T. \& Itoh, M.
Field-Induced Order-Disorder Transition in Antiferromagnetic BaCo$_2$V$_2$O$_8$ Driven by a Softening of Spinon Excitation.
{\it Phys. Rev. Lett.} {\bf 99}, 087602 (2007).

\bibitem{grenier2015} Grenier, B., Petit, S., Simonet, V., Can\'evet, E., Regnault, L.-P., Raymond, S., Canals, B., Berthier, C. \& Lejay, P.
Longitudinal and transverse Zeeman ladders in the Ising-like chain antiferromagnet BaCo$_2$V$_2$O$_8$.
{\it Phys. Rev. Lett.} {\bf 114}, 017201 (2015);
ibid.
Erratum: Longitudinal and transverse Zeeman ladders in the Ising-like chain antiferromagnet BaCo$_2$V$_2$O$_8$.
{\it Phys. Rev. Lett.} {\bf 115}, 119902 (2015).

\bibitem{ishimura1980} Ishimura N. \& Shiba, H.
Dynamical Correlation Functions of One-Dimensional Anisotropic Heisenberg Model with Spin l/2.
{\it Prog. Theor. Phys.} {\bf 63}, 743 (1980).

\bibitem{wang2015} Wang, Zhe M. Schmidt, M. Bera, A. K. Islam, A. T. M. N. Lake, B. Loidl, A. and Deisenhofer, J.
Spinon confinement in the one-dimensional Ising-like antiferromagnet SrCo$_2$V$_2$O$_8$.
{\it Phys. Rev. B} {\bf 91}, 140404(R) (2015).

\bibitem{wang2016} Wang, Zhe Wu, J., Xu, S. Yang, W. Wu, C. Bera, A. K. Islam, A. T. M. N. Lake, B. Kamenskyi, D. Gogoi, P. Engelkamp, H. Wang, N. Deisenhofer, J. and Loidl A.
From confined spinons to emergent fermions: Observation of elementary magnetic excitations in a transverse-field Ising chain.
{\it Phys. Rev. B} {\bf 94}, 125130 (2016).

\bibitem{bera2017} Bera, A. K. Lake, B. Essler,F. H. L. Vanderstraeten, L. Hubig, C. Schollw\"ock, U. Islam, A. T. M. N. Schneidewind,  A. and  Quintero-Castro D. L.
Spinon confinement in a quasi-one-dimensional anisotropic Heisenberg magnet.
{\it Phys. Rev. B} {\bf 96}, 054423 (2017).

\bibitem{sato2004} Sato, M., Oshikawa, M.
Coupled $S=1/2$ Heisenberg antiferromagnetic chains in an effective staggered field.
{\it Phys. Rev. B} {\bf 69}, 054406 (2004).

\bibitem{okutani2015} Okutani, A., Kimura, S., Takeuchi, T. \& Hagiwara, M.
High-Field Multi-Frequency ESR in the Quasi-1D $=1/2$ Ising-Like Antiferromagnet BaCo$_2$V$_2$O$_8$ in a Transverse Field.
{\it Appl. Magn. Reson.} {\bf 46}, 1003 (2015).

\bibitem{vidal2007} Vidal, G.
Classical Simulation of Infinite-Size Quantum Lattice Systems in One Spatial Dimension.
{\it Phys. Rev. Lett.} {\bf 98}, 070201 (2007).

\bibitem{phien2012} Phien, H. N., Vidal, G., \& McCulloch, I. P.
Infinite boundary conditions for matrix product state calculations.
{\it Phys. Rev. B} {\bf 86}, 245107 (2012).

\bibitem{AffleckOshikawa}
Affleck I. and Oshikawa M.
Field-induced gap in Cu benzoate and other $S=\frac{1}{2}$ antiferromagnetic chains.
{\it Phys. Rev. B} {\bf 60}, 1038 (1999).

\bibitem{berg_string}
Berg E., Dalla Torre E., Giamarchi T., Altman E.
Rise and fall of hidden string order of lattice bosons.
{\it Phys. Rev. B} {\bf 77}, 245119 (2008).

\bibitem{endres_string_cold}
Endres, M., Cheneau, M., Fukuhara, T., Weitenberg, C., Schauss, P., Gross, C., Mazza, L., Banuls, M.C., Pollet, L., Bloch, I., Kuhr, S.
Observation of Correlated Particle-Hole Pairs and String Order in Low-Dimensional Mott Insulators.
{\it Science} {\bf 334} 200 (2011).

\bibitem{Tsvelik2012} Tsvelik, A. M. \&Kuklov, A. B.
Parafermion excitations in a superfluid of quasi-molecular chains.
New J. Phys. {\bf 14}, 115033 (2012).

\bibitem{lejay2011} Lejay, P., Can\'evet, E., Srivastava, S. K., Grenier, B., Klanj\v{s}ek, M. \& Berthier, C.
Crystal growth and magnetic property of MCo$_2$V$_2$O$_8$ (M = Sr and Ba).
{\it J. Cryst. Growth} {\bf 317}, 128 (2011).

\bibitem{boehm2015} Boehm, M. , Steffens, P. , Kulda, J., Klicpera, M., Roux, S., Courtois, P., Svoboda, P., Saroun, J. \& Sechovsky, V.
ThALES--Three Axis Low Energy Spectroscopy for highly correlated electron systems.
{\it Neutron News}, {\bf 26:3}, 18-21 (2015), DOI: 0.1080/10448632.2015.1057050

\bibitem{Schmalzl2016} Schmalzl, K., Schmidt, W., Raymond,S., Feilbach, H., Mounier, C., Vettard, B. \& Br\"{u}ckel, T.
The upgrade of the cold neutron three-axis spectrometer IN12 at the ILL.
{\it Nuclear Instruments and Methods in Physics Research A} {\bf 819}, 89 (2016).
















\end{references}

\begin{references}

\bibitem{Wichmann1986}
Wichmann, R. \& M\"{u}ller-Buschbaum, Hk. Neue Verbindungen mit SrNi$_2$V$_2$O$_8$-Struktur: BaCo$_2$V$_2$O$_8$ und BaMg$_2$V$_2$O$_8$. {\it Z. Anorg. Allg. Chem.} {\bf 532}, 153 (1986).

\bibitem{Rodriguez-Carvajal93} Rodriguez-Carvajal, J. Recent advances in magnetic structure determination by neutron powder diffraction. {\it Physica B} {\bf 192}, 55 (1993).

\bibitem{canevet2013} Can\'evet, E., Grenier, B., Klanj\v{s}ek, M., Berthier, C., Horvati\'{c}, M., Simonet, V. \& Lejay, P. Field-induced magnetic behavior in quasi-one-dimensional Ising-like antiferromagnet BaCo$_2$V$_2$O$_8$: A single-crystal neutron diffraction study. {\it P. Phys. Rev. B} {\bf 87}, 054408 (2013).

\bibitem{kimura2008a} Kimura, S., Takeuchi, T., Okunishi, K., Hagiwara, M., He, Z., Kindo, K., Taniyama, T. \& Itoh, M. Novel Ordering of an $S=1/2$ Quasi-1d Ising-Like Antiferromagnet in Magnetic Field. {\it Phys. Rev. Lett.} {\bf 100}, 057202 (2008).

\bibitem{kimura2008b} Kimura, S., Matsuda, M., Masuda, T., Hondo, S., Kaneko, K., Metoki, N., Hagiwara, M., Takeuchi, T., Okunishi, K., He, Z., Kindo, K., Taniyama, T. \& Itoh, M. Longitudinal Spin Density Wave Order in a Quasi-1D Ising-like Quantum Antiferromagnet. {\it Phys. Rev. Lett.} {\bf 101}, 207201 (2008).

\bibitem{klanjsek2015} Klanj\v sek, M., Horvati\' c, M., Kr\" amer, S., Mukhopadhyay, S., Mayaffre, H., Berthier, C., Can\' evet, E., Grenier, B., Lejay, P.a \& Orignac, E. Giant magnetic-field dependence of the coupling between spin chains in BaCo$_2$V$_2$O$_8$. {\it Phys. Rev. B} {\bf 92} (2015) 060408(R).

\bibitem{grenier2015} Grenier, B., Petit, S., Simonet, V., Can\'evet, E., Regnault, L.-P., Raymond, S., Canals, B., Berthier, C. \& Lejay, P. Longitudinal and transverse Zeeman ladders in the Ising-like chain antiferromagnet BaCo$_2$V$_2$O$_8$. {\it Phys. Rev. Lett.} {\bf 114}, 017201 (2015); ibid. Erratum: Longitudinal and transverse Zeeman ladders in the Ising-like chain antiferromagnet BaCo$_2$V$_2$O$_8$. {\it Phys. Rev. Lett.} {\bf 115}, 119902 (2015).

\bibitem{kimura2006} Kimura, S., Yashiro, H., Hagiwara, M., Okunishi, K., Kindo, K., He, Z., Taniyama, T. \& Itoh, M. High field magnetism of the quasi one-dimensional anisotropic antiferromagnet BaCo$_2$V$_2$O$_8$. {\it J. Phys.: Conf. Ser.} {\bf 51}, 99 (2006).

\bibitem{coldea2010} Coldea, R., Tennant, D. A., Wheeler, E. M., Wawrzynska, E., Prabhakaran, D., Telling, M., Habicht, K., Smeibidl, P. \& Kiefer, K. Quantum Criticality in an Ising Chain: Experimental Evidence for Emergent E8 Symmetry. {\it Science} {\bf 327}, 177 (2010).

\bibitem{vidal2007}
Vidal, G.
Classical Simulation of Infinite-Size Quantum Lattice Systems in One Spatial Dimension. {\it Phys. Rev. Lett.} \textbf{98}, 070201 (2007).

\bibitem{phien2012}
Phien, H. N., Vidal, G. \& McCulloch, I. P.
Infinite boundary conditions for matrix product state calculations.
{\it Phys. Rev. B} \textbf{86}, 245107 (2012).

\bibitem{kimura2013}
Kimura, S., Okunishi, K., Hagiwara, M., Kindo, K., He, Z., Taniyama, T., Itoh, M., Koyama, K. \& Watanabe, K. Collapse of Magnetic Order of the Quasi One-Dimensional Ising-Like Antiferromagnet BaCo$_{2}$V$_{2}$O$_{8}$ in Transverse Fields. {\it J. Phys. Soc. Jpn.} \textbf{82}, 033706 (2013).

\bibitem{giamarchibook}
Giamarchi, T. Quantum Physics in One Dimension (Oxford University Press, Oxford, 2004).
\end{references}
\end{document}